\PassOptionsToPackage{usenames,dvipsnames}{xcolor}
\documentclass[a4paper,11pt]{article}
\usepackage[version=3]{mhchem}
\usepackage{pos}
\usepackage{amsmath,pdftexcmds}
\usepackage{booktabs}
\usepackage{multirow}
\usepackage{tabularx}
\usepackage{amsfonts}
\usepackage{mathtools}
\usepackage{amsbsy}
\usepackage[T1]{fontenc}
\usepackage{dsfont}
\usepackage{hyperref}

\usepackage{pgfplots}
\usepackage{pgfplotstable}
\usepackage{tikz}
\usepackage{tikz-3dplot}
\usetikzlibrary{trees}
\usetikzlibrary{decorations.markings}
\usetikzlibrary{calc}
\usetikzlibrary{plotmarks}
\usetikzlibrary{fpu,arrows} 
\usepgfplotslibrary{fillbetween}

\definecolor{DarkMidnightBlue}{rgb}{0.0, 0.04, 0.14}
\definecolor{Bittersweet}{rgb}{1.0, 0.44, 0.37}
\definecolor{Burgundy}{rgb}{0.5, 0.0, 0.13}
\definecolor{Caribbeangreen}{rgb}{0.0, 0.8, 0.6}
\definecolor{Lilla}{rgb}{0.71, 0.4, 0.82}
\definecolor{Hotmagenta}{rgb}{1.0, 0.11, 0.81}
\definecolor{Tangerine}{rgb}{0.95, 0.52, 0.0}

\renewcommand{\logo}{\relax}

\hypersetup{hidelinks,
backref=true,
pagebackref=true,
hyperindex=true,
breaklinks=true,
colorlinks=true,
linkcolor=blue, 
citecolor=blue, 
urlcolor=blue,
bookmarks=true,
bookmarksopen=false,
pdftitle={Title},
pdfauthor={Author}}

\DeclareFontFamily{U}{cbgreek}{}
\DeclareFontShape{U}{cbgreek}{m}{n}{
        <-6>    grmn0500
        <6-7>   grmn0600
        <7-8>   grmn0700
        <8-9>   grmn0800
        <9-10>  grmn0900
        <10-12> grmn1000
        <12-17> grmn1200
        <17->   grmn1728
      }{}
\DeclareFontShape{U}{cbgreek}{bx}{n}{
        <-6>    grxn0500
        <6-7>   grxn0600
        <7-8>   grxn0700
        <8-9>   grxn0800
        <9-10>  grxn0900
        <10-12> grxn1000
        <12-17> grxn1200
        <17->   grxn1728
      }{}

\DeclareRobustCommand{\digamma}{%
  \text{\usefont{U}{cbgreek}{\normalorbold}{n}\symbol{147}}%
}

\makeatletter
\newcommand{\normalorbold}{%
  \ifnum\pdf@strcmp{\math@version}{bold}=\z@ bx\else m\fi
}
\makeatother

\newcommand\mscriptsize[1]{\mbox{\scriptsize\ensuremath{#1}}}

\title{Spectrum and electromagnetic properties of \ce{^{24}Mg} in the Geometric $\alpha$-cluster Model with $\mathcal{D}_{4h}$ symmetry at leading order}
\ShortTitle{Spectrum and EM properties of \ce{^{24}Mg} in the G$\alpha$CM with $\mathcal{D}_{4h}$ symmetry at LO}

\author*[a,b]{Gianluca Stellin}
\author[c]{Karl-Heinz Speidel$^{\dagger,}$\hspace{-2mm}\note[$\dagger$]{Deceased on June 19, 2023.}}

\affiliation[a]{DRF/IRFU/DPhN/LENA, ESNT, CEA Paris-Saclay, \\
91191 Gif-sur-Yvette, France}
\affiliation[b]{IJCLab, CNRS-In2p3, \textit{Université Paris-Saclay}, 91405 Orsay , France}

\affiliation[c]{Helmholtz Institut f\"ur Strahlen- und Kernphysik, Universit\"at Bonn,\\
Nu\ss{}allee 14-16, 53115 Bonn, Germany}

\emailAdd{gianluca.stellin@ijclab.in2p3.fr}

\abstract{
The relevance of the point-symmetry group $\mathcal{D}_{4h}$ for the prediction of spectrum and electromagnetic properties of the \ce{^{24}Mg} nucleus is discussed in the 
framework of the geometric $\alpha$-cluster model at leading order. The latter represents a macroscopic $\alpha$-cluster framework, in which nuclear excitations are described in terms of
rotations and vibrations of \ce{^4He} clusters about their equilibrium positions, at the vertices of a square bipyramid. The finite group associated with the latter regulates the composition
of the rotational bands as well as the transitions between the energy levels, by means of additional selection rules, of molecular nature. A sample of reduced electric multipole transition probabilities 
of intraband nature is provided.
}

\FullConference{
\begin{center}
\begin{minipage}{0.30\columnwidth} 
\includegraphics[width=1.00\columnwidth]{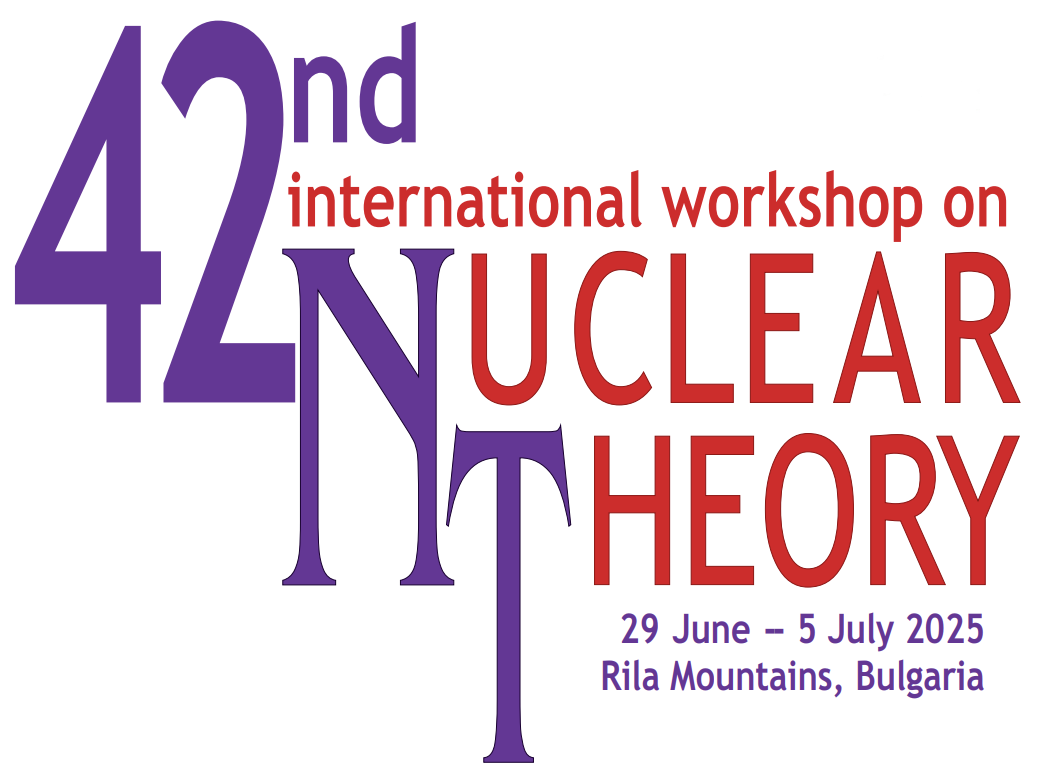} 
\end{minipage}\hspace{2mm}
\begin{minipage}{0.40\columnwidth}
\textbf{42th International Workshop}\\
\textbf{on Nuclear Theory}\\
Bulgarian Academy of Sciences,\\
June 30 - July 4, 2025, \\
Hotel Lion****,\\
2010 Borovec, \\
Bulgaria
\end{minipage}
\end{center}
}

\begin{document}
\maketitle

\section{Introduction}\label{sec:introduction}

The spectrum of the \ce{^{24}Mg} nucleus and its structural properties have been investigated far and wide, namely in relation with cluster configurations. An early calculation of the electromagnetic (EM) transition strengths has been obtained through a microscopic rotational model \cite{Hor72}, in which the intrinsic states are given by Slater determinants of the single-particle eigenstates of a Nilsson-like Hamiltonian, with the spin-orbit interaction. 

In the literature, the intrinsic deformation of this open-shell isotope has been modeled by assuming an internal partition of the nucleons in two or more aggregates of nucleons, such as the $2\alpha$ + \ce{^{16}O} \cite{KaH79}, $\alpha$ + \ce{^{20}Ne} \cite{KaH79} and \ce{^{12}C} + \ce{^{12}C} \cite{BHM90,CLW93,CRJ23} cluster configurations. The results in the Buck-Dover-Vary cluster (BDVC) model \cite{BHM90} and in the semimicroscopic algebraic cluster model (SACM) \cite{CLW93}, recommend to prefer open-shell clusters as \ce{^{12}C}, since the latter can incorporate part of the deformation of the nucleus and generate a larger amount of excited states.

More in recent times, the antisymmetrized molecular dynamics (AMD) me\-thod has been applied to this nucleus, mostly with the purpose of analysing the intrinsic matter density distribution. So far, the latter has been studied for states lying in the two lowest $K^{\pi}=0^+$ rotational bands and the lowest $2^+$ band \cite{KFT12,KiC14}, also in relation with isoscalar E0 \cite{ChK15} and E2 transitions \cite{KYI12}. In the latter work, the AMD approach has been combined with the generator coordinate method (GCM) with the purpose of processing a large number of basis wavefunctions \cite{ChK15}. The same method has been adopted for the structural analysis of the negative-parity states in the lowest $K^{\pi}=0^-$ and $1^-$ bands in Ref.~\cite{ChK20}, whereas a more complete survey has been provided in Ref.~\cite{KEO21}, encompassing also the form factors. In the latter, AMD has been exploited in combination with the variation-after-parity-and-total-angular-momentum projection (VAP) method for the structure properties and with the microscopic coupled-channel approach (MCC) for the study of proton and $\alpha$-particle scattering off the \ce{^{24}Mg} nucleus.   

In terms of \textit{ab-initio} approaches, the ground state features have been analysed by nuclear lattice effective field theory (NLEFT) \cite{EBM24}, adopting realistic nuclear forces drawn from chiral effective field theory ($\chi$EFT) at $N^3LO$. A NLEFT-based analysis of the structure of the excited states as in Ref.~\cite{SEL23} is envisaged \cite{SSL26}. Additionally, isoscalar and isovector resonances have been investigated by means of the \textit{ab-initio} projected generator coordinate method (PGCM) \cite{PDE24}.

Furthermore, the recent investigations on spectrum and electromagnetic properties of ${}^{20}\mathrm{Ne}$ \cite{BiI21-01,Gan25} together with the application of the cluster shell model (CSM) to ${}^{21}\mathrm{Ne}$ and ${}^{21}\mathrm{Na}$ \cite{BiI21-02} have reignited the interest in $\alpha$-clustering on $sd$-shell nuclei. Besides, a recent study on \ce{^{12}C} \cite{KiT24} provides additional support to the finding that not only the states lying in the Hoyle band exhibit developed $\alpha$-cluster features, but also for the levels in the ground state band the ${}^{4}\mathrm{He}$ clusters represent relevant nuclear subunits. 

Historically \cite{Wef37,Whe37,HaT38}, macroscopic $\alpha$-cluster models, \textit{i.e.} phenomenological approaches employing the ${}^{4}\mathrm{He}$ clusters as the sole degrees of freedom have been applied to light $\alpha$-conjugate nuclei, mainly ${}^{12}\mathrm{C}$ \cite{SFV16,GlG56,BiI02,Jen16,PoC79} and ${}^{16}\mathrm{O}$ \cite{Den40,BiI17,For24,RBL06,BiI14}. Resting on the assumption that the oscillation period of the ${}^{4}\mathrm{He}$ clusters is smaller or comparable with the diffusion time of a nucleon inside a cluster, nuclear excitations can be described in terms of rotating and vibrating of $\alpha$-clusters around their molecular equilibrium configurations \cite{Whe37,SFV16}. The associated finite discrete symmetries are referred to as molecular or \textit{exotic} nuclear symmetries \cite{DCD18,DeD24} and underlie also the more recent macroscopic $\alpha$-cluster approaches, such as the \textit{algebraic cluster model} (ACM) \cite{BiI00,BiI20} and the \textit{quantum graph model} (QGM) \cite{Hal17,HaR18,Raw18}. The choice of the $\alpha$-cluster ansatz is dictated by the number of bonds, $n_b$, between nearest-neighbour \ce{^4He} clusters; the spin and parity of the levels assigned in the literature to the \textit{ground-state} (\textit{g.s.}) rotational band; the magnitude and the sign of the measured electric quadrupole moment of the nucleus in a state of the \textit{g.s.} band. 

Specifically, for \ce{^{24}Mg}, $\mathcal{D}_{4h}$ symmetry has quite firm roots in the literature \cite{Bou62,NTD66}. The same point group appears in early applications \cite{BFW70} of the Bloch-Brink alpha-cluster model (BBACM) \cite{Bri65}. In fact, by starting from a rectangular bypiramidal $\alpha$-particle structure with Volkov or Brink-Boeker interactions, one finds that the energy eigenvalue of the ground state of the nucleus is minimized by the square bipyramidal configuration \cite{BSS80}. Nevertheless, due to the difficulty in reproducing the experimental transition form factor to the $2_1^+$ state, $\alpha$-cluster equilibrium configurations with different invariances, such as the $\mathcal{D}_{2h}$ symmetry associated with a bitetrahedron, have been proposed \cite{HWD71,Hau71}. Nevertheless, the quite good agreement obtained with a $\mathcal{D}_{4h}$-symmetric elastic form factor \cite{Ste26} (cf. Sec.~\ref{sec:unexcited_A1g_band}), the individuation of nine candidates for the lowest singly-excited rotational bands \cite{StS24} (cf. Secs.~\ref{sec:excited_A1g_I_band}-\ref{sec:excited_Eu_II_band}) as well as the good accuracy measured reduced E0 and E2 transition probabilities recapitulated in Ref.~\cite{Ste26} substantiate the adoption of the square-bipyramidal equilibrium configuration. Finally, the flexibility of the formalism of the geometric $\alpha$-cluster model (G$\alpha$CM) allows for a systematic improvement in the prediction of nuclear observables, including the transition form factors, by considering the coupling between rotational and vibrational motion of the ${}^{4}\mathrm{He}$ clusters.

\begin{figure}[ht!]
    \centering
    \begin{minipage}{0.49\columnwidth}
    \includegraphics[width=0.99\columnwidth]{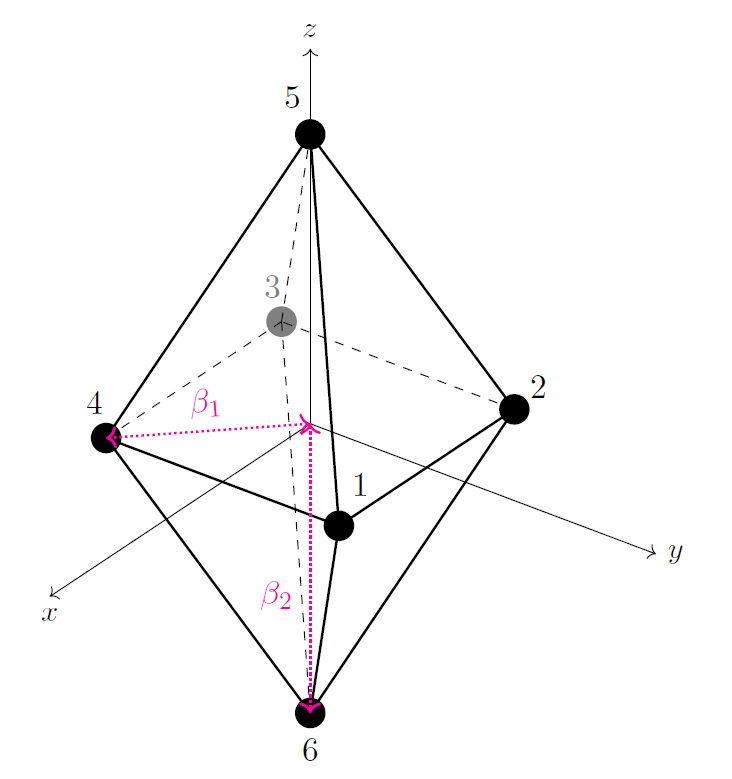}
        \label{fig:MicroscopicD4hStructure}
        \end{minipage}
    \begin{minipage}{0.49\columnwidth}
    \hspace{4mm}
        \includegraphics[width=0.82\columnwidth]{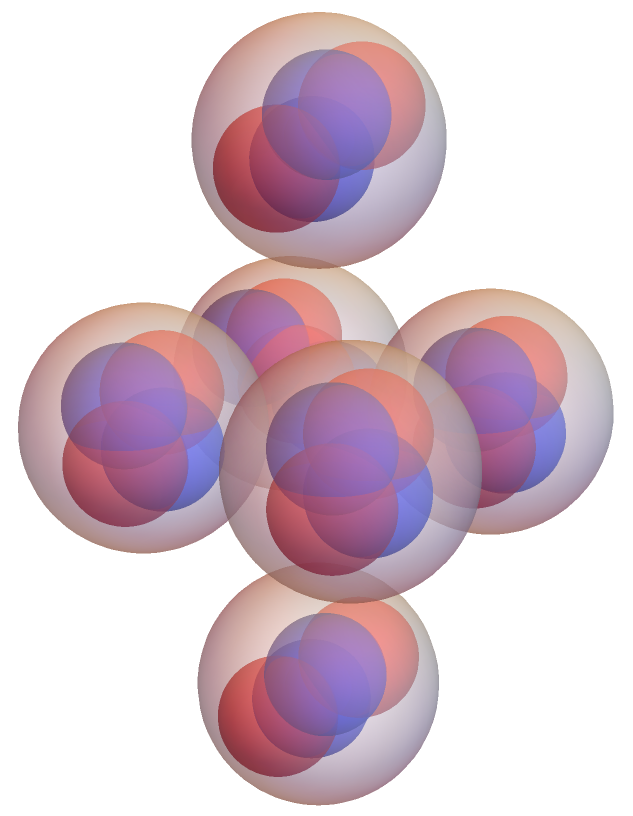}
    \label{fig:MacroscopicD4hStructure}
    \end{minipage}
    \caption{Equilibrium $\alpha$-cluster configuration of ${}^{24}\mathrm{Mg}$ in the intrinsic reference frame (left) with the underlying microscopic structure in terms for protons (red) and neutrons (blue) with realistic charge radii (right). The structure parameters $(\beta_1,\beta_2)$, highlighted in red, are evaluated at $(2.380, 3.857)$~fm, corresponding to a prolate shape, consistently with the measured charge radius of the $0_1^+$ state and the electric quadrupole moment of the $2_1^+$ state. The charge distribution of the $\alpha$-particles is assumed to be pointlike.}
    \label{fig:24Mg_SquareBipyramid_D4hStructure}
\end{figure}

Effectively, the \ce{^{24}Mg} nucleus has been subject to a number of ab-initio and antisymmetrized molecular dynamics calculations. By means of ab-initio approaches, the ground state properties have been analysed by different approaches implementing realistic nuclear interactions from Chiral effective field theory ($\chi$EFT), notably via nuclear lattice effective field theory (NLEFT) \cite{EBM24}. Recently, certain resonances have been investigated through the \textit{ab-initio} projected generator coordinate method (PGCM) \cite{PDE24},

\section{Formalism}\label{sec:formalism}

The theoretical framework is provided by the \textit{geometric $\alpha$-cluster model} (G$\alpha$CM) \cite{StS24,Ste26}, describing the nucleus in terms of 6 $\alpha$-particles rotating and vibrating about their equilibrium positions \cite{Ste15}, located at the vertices of a square bipyramid (cf. Fig.~\ref{fig:24Mg_SquareBipyramid_D4hStructure}). 

The most general Hamiltonian of a system of six harmonically vibrating and rotating clusters coincides with Watson's Hamiltonian \cite{Wat68}, in which the two collective motions are completely coupled. Nonetheless, if the departures of the $\alpha$-clusters, $\Delta\alpha_i$\footnote{The notation of Refs.~\cite{BuJ04,Ste15,StS24} is adopted: $\alpha,\beta,\gamma \ldots = x,y,z$ ($A,B,C, \ldots =\xi, \eta, \zeta$) denote the Cartesian components of tensors defined in the body-fixed or \textit{intrinsic} (laboratory) frame.}, from their equilibrium positions in the body-fixed reference frame, $\alpha_i^e$, are small with respect to the size of the system, characterized by the parameters $(\beta_1,\beta_2)$ in Fig.~\ref{fig:24Mg_SquareBipyramid_D4hStructure}, the rotation-vibration coupling contributions can be built up on top of the rigid-rotor Hamiltonian. The rotation-vibration coupling terms are obtained from the expansion of the \textit{effective} reciprocal inertia tensor,
\begin{equation}
\mu_{\alpha\beta}^{-1} = I_{\alpha\beta} - \sum_{k=1}^{3N-6}\left( \sum_{j=1}^{3N-6}\zeta_{jk}^{\alpha}Q_j\sum_{l=1}^{3N-6}\zeta_{lk}^{\beta}Q_l\right)~,
\end{equation}
into power series of the elements $I_{\alpha\beta}$ along the directions $\alpha$ and $\beta$ of the inertia tensor. By splitting the latter into the \textit{static} contribution, $I_{\alpha\beta}^{\mathrm{stat}}$, depending solely on the equilibrium $\alpha$-cluster positions, and the \textit{dynamic} contribution, $I_{\alpha\beta}^{\mathrm{dyn}}$, depending quadratically on the \textit{normal coordinates} of vibration, $Q_i$ with $i=1, \ldots 12$, one can write
\begin{eqnarray}
\boldsymbol{\mu}  = (\mathbf{I}^{\mathrm{stat}} + \mathbf{I}^{\mathrm{dyn},\zeta})^{-1} & = & \left(\mathds{1} - {\mathbf{I}^{\mathrm{stat}}}^{-1}\mathbf{I}^{\mathrm{dyn},\zeta} \right. \\  &+& \left.  {\mathbf{I}^{\mathrm{stat}}}^{-1}  \mathbf{I}^{\mathrm{dyn},\zeta}{\mathbf{I}^{\mathrm{stat}}}^{-1}\mathbf{I}^{\mathrm{dyn},\zeta} + \ldots\right)  {\mathbf{I}^{\mathrm{stat}}}^{-1} \nonumber
\label{eqn:EffectiveInertiaTensorExpansion}
\end{eqnarray}
where $\zeta_{jk}^{\gamma} \in \mathbb{R}$ are c-numbers, defined in Refs.~\cite{Ste15,StS24}.  In order to understand the significance of the normal coordinates graphically, it is convenient to express them in terms of the displacement coordinates $\Delta\alpha_i$ in the body-fixed frame. 
Under the constraint of normalization to $\sqrt{m}$ \cite{StS24}, the normal coordinates of the six non-degenerate modes take the form of Eqs. (13a)-(14f) in Ref.~\cite{StS24}, with $Q_1$ and $Q_2$ in Eqs.~(13a)-(13b) replaced by the following linear combinations,
\begin{equation}
Q_1' = \frac{1}{\sqrt{2}}Q_1 + \frac{1}{\sqrt{2}}Q_2\label{eqn:normalccordinatenewQ1}~,
\end{equation}
\begin{equation}
Q_2' = \frac{1}{\sqrt{2}}Q_1 - \frac{1}{\sqrt{2}}Q_2\label{eqn:normalccordinatenewQ2}~.
\end{equation}
The linear combnations in Eqs.~\eqref{eqn:normalccordinatenewQ1}-\eqref{eqn:normalccordinatenewQ2} are legitimate, since the original modes in Eqs. (13a)-(13b) in Ref.~\cite{StS24} $Q_1$ and $Q_2$ transform according to the same irreducible representation of the symmetry group associated with the equilibrium configuration of the $\alpha$-particles,  $\mathcal{D}_{4h}$. The latter consists of 16 elements, subdivided into 10 conjugacy classes containing proper and improper rotations (cf. Ref.~\cite{StS24} for the character table). The ensuing 10 irreducible representations \guillemotleft irreps \guillemotright~are denoted as $A_{1g}$, $A_{2g}$, $A_{1u}$, $A_{2u}$ (one-dimensional and even under the rotations about the principal symmetry axis), $B_{1g}$, $B_{2g}$, $B_{1u}$, $B_{2u}$ (one-dimensional and odd under the rotations about the principal symmetry axis), $E_g$ and $E_u$ (two-dimensional).  The $g$ ($u$) superscript is assigned to representations with positive (negative) parity, consistently with Mulliken's notation \cite{Car97}.

\begin{figure}[ht!]
    \centering
    \includegraphics[width=0.99\columnwidth]{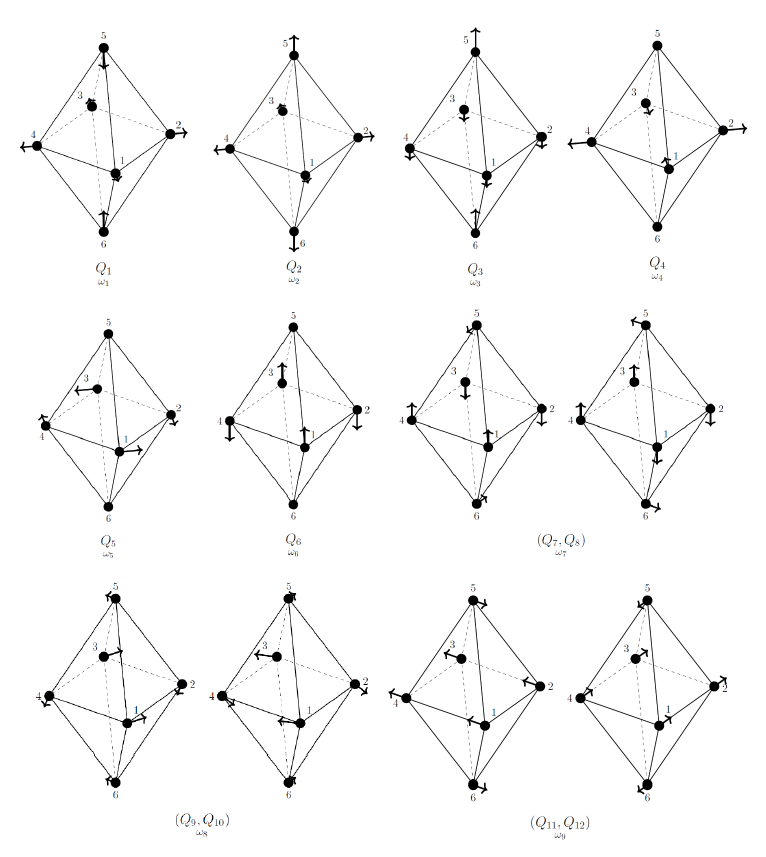}
    \caption{
    Normal vibrations of a square bipyramidal configuration with $\mathcal{D}_{4h}$ symmetry. The oriented segments with arrows denote the displacements of the $\alpha$-clusters with respect to their equilibrium positions. The 'prime' superscripts in the modes with frequencies $\omega_1$ and $\omega_2$ are suppressed.
    }
    \label{fig:24Mg_D4h_NormalModes}
\end{figure}

In particular, the coordinate $Q_1'$ represents an asymmetric \textit{stretching} mode and transforms according to the $A_{1g}$ representation of $\mathcal{D}_{4h}$, whereas $Q_2'$ is a symmetric \textit{stretching} or~\guillemotleft breathing\guillemotright~mode, with the same transformation properties under the operation of the equilibrium symmetry group (cf. Fig.~\ref{fig:24Mg_D4h_NormalModes}) . Conversely, the $Q_3$ mode denotes a symmetric \textit{wagging} mode (irrep $A_{2u}$), the $Q_4$ an asymmetric \textit{stretching} mode (irrep $B_{1g}$), the $Q_5$ a symmetric \textit{scissoring} mode (irrep $B_{2g}$) and the $Q_6$ a symmetric \textit{twisting} mode \cite{Ste26} (irrep $B_{2u}$).

As for the two $A_{1g}$ modes, it is possible to write alternative normal coordinates for the two $E_u$ modes, by means of linear combinations. However, in order to facilitate the construction of eigenstates states with well-defined symmetry properties under the operations of $\mathcal{D}_{4h}$, it is preferable to select pairs of normal coordinates, $(Q_9',Q_{10}')$ and $(Q_{11}',Q_{12}')$, transforming separately under the operations of the equilibrium symmetry group. More in general, for the doubly-degenerate modes, the expressions of the $Q_i$'s in terms of the displacement coordinates in the intrinsic frame are not unique, since valid normal coordinates can be obtained by taking linear combinations of the original coordinates in the respective pairs, $(Q_7,Q_8)$, $(Q_9,Q_{10})$ and $(Q_{11},Q_{12})$. In Fig.~\ref{fig:24Mg_D4h_NormalModes}, the coordinates $(Q_7,Q_8)$ represent an asymmetric \textit{twisting} mode (irrep $E_g$), the $(Q_9,Q_{10})$ an asymmetric \textit{scissoring} mode (irrep $E_u$) and the $(Q_{11},Q_{12})$ an asymmetric \textit{rocking} mode \cite{Ste26} (irrep $E_u$). By defining suitable matrix operators for the $16$ elements of $\mathcal{D}_{4h}$ \cite{Ste26} for the normal coordinates, one obtains the transformation properties recapitulated in Sec.~3 of Ref.~\cite{StS24}. 

In Eq.~\eqref{eqn:EffectiveInertiaTensorExpansion}, $I_{\alpha\beta}^{\mathrm{dyn, \zeta}}$ denote the components of the \textit{effective} dynamic inertia tensor,
\begin{equation}
I_{\alpha\beta}^{\mathrm{dyn, \zeta}} \equiv I_{\alpha\beta}^{\mathrm{dyn}} - \sum_{k=1}^{12}\left( \sum_{j=1}^{12}\zeta_{jk}^{\alpha}Q_j\sum_{l=1}^{12}\zeta_{lk}^{\beta}Q_l\right)~.
\end{equation}
The contributions in Eq.~(\ref{eqn:EffectiveInertiaTensorExpansion}), together with the \emph{vibrational} angular momentum, 
\begin{equation}
p_{\alpha}=-i\hbar\sum_{jk=1}^{12}\zeta_{jk}^{\alpha}Q_j\frac{\partial}{\partial Q_k}~,
\end{equation}
are grouped consistently with a power-counting scheme, which permits to construct a tower of systematically-improved Hamiltonians, $H_{LO}$, $H_{NLO}$, $H_{N^2LO} \ldots$ \cite{Ste26} corresponding to higher-order rotation-vibration correlations built on top of the exact Hamiltonian \cite{Wat68}. 

A survey of systematic approximation schemes for Watson's Hamiltonian in molecular physics is provided in Ref.~\cite{PBB15}. 
The various interaction terms coupling rotations with vibrations might be treated in perturbation theory, built on top of the reference LO Hamiltonian. In general, the corrections do not conserve $\mathcal{D}_{4h}$ symmetry and induce triaxiality, although to different extent, depending on the truncation scheme and the excited vibrational quanta.

In particular, the LO Hamiltonian coincides with the well-known \textit{rigid-rotor} limit,
\begin{eqnarray}
H_{LO} = \frac{J^2}{2I_{xx}^{\mathrm{stat}}}&-&\frac{J_z^2}{2}\left(\frac{1}{I_{xx}^{\mathrm{stat}}}-\frac{1}{I_{zz}^{\mathrm{stat}}}\right) - \frac{\hbar^2}{2}\sum_{j=1}^{12}\frac{\partial^2}{\partial Q_j^2} \nonumber \\ &+& \frac{1}{2}\sum_{j=1}^{12}\lambda_j^2 Q_j^2  - \frac{\hbar^2}{8}\sum_{\alpha}\mu_{\alpha\alpha}^{\mathrm{stat}}~,
\label{eqn:RigRotHamiltonian}
\end{eqnarray}
where $(\lambda_1,~\lambda_2~\ldots~\lambda_{12})$ $ \equiv (\omega_1,\omega_2,\omega_3, \omega_4, \omega_5, \omega_6,\omega_7,\omega_7,\omega_8,\omega_8,\omega_9,\omega_9)$ are the \textit{frequencies} associated with the normal modes with coordinates $Q_1,~Q_2,~\ldots~Q_{12}$ respectively
and 
\begin{subequations}
\begin{equation}
I_{xx}^{\mathrm{stat}}  =  I_{yy}^{\mathrm{stat}} = 2m (\beta_1^2 + \beta_2^2)~, 
\label{eqn:StatInertiaTensor_pl}
\end{equation}
\begin{equation}
I_{zz}^{\mathrm{stat}}  = 4m\beta_1^2~,
\label{eqn:StatInertiaTensor_ax}
\end{equation}
\end{subequations}
are the only non-vanishing components of the static inertia tensor, as the axes of the body-fixed frame are parallel to the principal axes of inertia of the square bipyramid and $m \approx 3727.4~\mathrm{MeV}$ is the $\alpha$-particle mass. In addition, the components of the \emph{rotational} angular momentum operator along the body-fixed axes are given by Eqs. (7a)-(7c) in Ref.~\cite{StS24} in terms of the Euler angles $(\chi,\theta,\varphi)$ in the active picture \cite{VdW01}. The invariances of $H_{LO}$ are larger than the ones of Watson's Hamiltonian \cite{Wat68}, since rotations and vibrations are decoupled and the system behaves as a symmetric top (cf. Eq.~(\ref{eqn:StatInertiaTensor_pl})-(\ref{eqn:StatInertiaTensor_ax})).  
Although the EM observables in this proceeding are calculated at LO, i.e. the uncorrelated limit, it is of interest to display the NLO Hamiltonian $H_{NLO} \equiv H_{LO}+H_{NLO}^{\mathrm{corr}}$, where
\begin{eqnarray}
H_{NLO}^{\mathrm{corr}} & \equiv & - \frac{1}{2}\sum_{\alpha\beta} J_{\alpha}({I_{\alpha\beta}^{\mathrm{stat}}}^{-1} I_{\beta\gamma}^{\mathrm{dyn},\zeta}{I_{\gamma\delta}^{\mathrm{stat}}}^{-1})J_{\delta} - \sum_{\alpha\beta}J_{\alpha}{I_{\alpha\beta}^{\mathrm{stat}}}^{-1}p_{\beta} \nonumber \\ & + & \frac{1}{2}\sum_{\alpha\beta} p_{\alpha}({I_{\alpha\beta}^{\mathrm{stat}}}^{-1})p_{\beta} + \frac{\hbar^2}{8}\left({I_{\alpha\beta}^{\mathrm{stat}}}^{-1} I_{\beta\gamma}^{\mathrm{dyn},\zeta} {I_{\gamma\alpha}^{\mathrm{stat}}}^{-1}\right) ~,\label{eqn:H_GaCM_NLO}
\end{eqnarray}
which incorporates the lowest-order contributions from the vibrational angular momentum.

Since $H_{LO}$ commutes with the components of the angular momentum operator in the laboratory frame, $J_{\xi}$, $J_{\eta}$ and $J_{\zeta}$, as well as with $J_{z}$ in Eq.~(7c) of Ref.~\cite{StS24}, the eigenvalues of Eq.~(\ref{eqn:RigRotHamiltonian}) admit an expression in closed form,
\begin{eqnarray}
E_{LO}(J,K,[\mathfrak{n}]) &=& \frac{\hbar^2}{2 I_{xx}^{\mathrm{stat}}}[J(J+1) -  K^2] + \frac{\hbar^2 K^2}{2 I_{zz}^{\mathrm{stat}}} - \frac{\hbar^2}{8}\sum_{\alpha}{I_{\alpha\alpha}^{\mathrm{stat}}}^{-1} \nonumber \\ &+& \sum_{i=1}^6\hbar\omega_i(\mathfrak{n}_i+ \frac{1}{2}) + \sum_{i=7}^9\hbar\omega_i(\mathfrak{n}_i+ 1) ~,
\label{eqn:HLOeigenvalues}
\end{eqnarray}
where $\hbar K$ ($\hbar M$) is the angular momentum projection along the intrinsic (labora\-tory-fixed) $z$-axis and $\hbar^2 J(J+1)$ is the eigenvalue of the quadratic Casimir operator of $\mathfrak{so}(3)$, $J^2$. In Eq.~(\ref{eqn:HLOeigenvalues}), the number of vibrational quanta or \guillemotleft phonons\guillemotright~for the normal modes $\mathfrak{n}_i$ are vectorized as $[\mathfrak{n}]$, where $i=1,2\ldots 9$, because the modes $i=7,8$ and $9$ are two-dimensional. 

Since rotations and vibrations are decoupled, the eigenstates of $H_{LO}$ can be factorized into a vibrational part, $\psi_V$, and a rotational part, $\psi_{R}$. Nonetheless, due to the presence of a combination of an axial symmetry and a point-group symmetry, $\mathcal{D}_{4h}$, the factorization of the eigenstates of $H_{LO}$ into rotational and vibrational states (cf. Sec.~4-2c of Ref.~\cite{BoM75-II}) is no longer valid.

By denoting the vibrational quanta with the frequencies $\lambda_i$, \textit{i.e.} $\nu_i = \mathfrak{n}_i$ with $i=1,2,\ldots 6$ for the non-degenerate modes and $\nu_7+\nu_8 = \mathfrak{n}_7$, $\nu_9+\nu_{10} = \mathfrak{n}_8$ and $ \nu_{11}+\nu_{12} = \mathfrak{n}_{9}$ for the doubly-degenerate ones, the vibrational part of the eigenfunctions can be succinctly expressed as
\begin{equation}
 \psi_V(Q_1, Q_2 \ldots Q_{3N-6}) =\prod_{i=1}^{12}  \Phi_{{\nu}_i}(Q_i)~,
 \label{eqn:vibrationalstatesLO}
\end{equation}
where $\Phi_{{\nu}_i}$ are 1D harmonic-oscillator eigenfunctions with frequency $\omega_i$,
\begin{equation}
\Phi_{{\nu}_i}(Q_i) = \frac{1}{\sqrt{\nu_i!~2^{\nu_i}}} \left(\frac{\omega_i}{\pi\hbar}\right)^{1/4} H_{\nu_i}({\textstyle\tiny{ \sqrt{\frac{\omega_i}{\hbar}} }}Q_i)~e^{-\frac{\omega_i}{2\hbar}Q_i^2}~,
\label{eqn:HOvibreigenfunctions}
\end{equation}
and $H_{\nu_i}$ is a Hermite polynomial of degree $\nu_i$. The HO eigenstates of the doubly-degenerate modes can be equivalently recast into the polar basis \cite{KBH14}, as in Ref.~\cite{Ste15}. 

Moreover, the vibratonal eigenfunction, $\psi_V$, corresponding to zero vibration quanta, $\mathfrak{n} = \mathbf{0}$, transforms as the trivial irreducible representation, $A_{1g}$. Besides, the harmonic oscillator states with a single phonon in the non-degenerate mode $\omega_i$ transform according to the same irreducible representation as the normal coordinate $Q_i$. Analogously, states with $\mathfrak{n}_7$, $\mathfrak{n}_8$ or $\mathfrak{n}_9 =1$ transform according to the $E_g$, $E_u$ and $E_u$ representations respectively. 

Beyond one quantum of excitation, the eigenfunctions with an odd number of oscillator quanta in the 1D mode $\omega_i$ behave as the coordinate $Q_i$. If the number of quanta in a non-degenerate mode is even, the corresponding harmonic oscillator eigenfunction transforms as the $A_{1g}$ representation. Conversely, the eigenfunctions corresponding to multiple vibration quanta $\mathfrak{n}_i$ in the doubly-degenerate mode with irreducible representation $\Gamma_2=E_g$ or $E_u$ transform according to a reducible representation whose characters are obtained from the symmetric $\mathfrak{n}_i$-th power of the irrep $\Gamma_2$ (cf. Tabs.~2-3 in Ref.~\cite{StS24}).

Because $SO(3) \supset SO(2)$ constitute dynamical symmetries of $H_{LO}$, referring to 3D rotations in the laboratory-fixed frame and 2D rotations along the z-axis of the intrinsic frame, 
the rotational states can be labeled with the eigenvalues of the Casimir operators of the algebras of the former two groups, $\mathfrak{so}(3)$ and $\mathfrak{so}(2)$ \cite{GrM96}, 
\begin{equation}
\psi_{R}(\chi,\theta,\varphi) \equiv \langle \chi, \theta, \varphi | J, M, K \rangle = \sqrt{\frac{(2J+1)}{8\pi^2}}D_{KM}^{J*}(\chi,\theta,\varphi)~,
\label{eqn:rotstatesRR}
\end{equation}
where $D_{KM}^{J*}(\chi,\theta,\varphi)$ are Wigner D-matrices in the active picture \cite{VdW01}. For the rotational wavefunctions in Eq.~\eqref{eqn:rotstatesRR}, direct inspection of the transformation properties of $\psi_R$ with different angular momenta as in Ref.~\cite{Ste15}, delivers the results in Tab.~4 of Ref.~\cite{StS24}. The fact that the $\alpha$-structure possesses at least an invariance under a rotation of angle $\pi$ about an axis orthogonal to the symmetry axis (\textit{e.g.} the intrinsic $y$-axis), in fact, has been exploited (cf. Sec. 4-2c of Ref.~\cite{BoM75-II}). As a consequence, all the operations of $\mathcal{D}_{4h}$ behave on the Wigner D-matrices as proper rotations (cf. Sec. 4-2d of Ref.~\cite{BoM75-II}) and $\psi_R$ has positive parity, regardless of $J$ and $K$. 

As shown in Ref.~\cite{StS24}, the restrictions imposed on the LO eigenfunctions by the $\mathcal{D}_{4h}$ group result into
\begin{eqnarray}
\Psi_{RV}^{\mathcal{D}_{4h}}(\mathbf{Q}, \boldsymbol{\Omega})  & = & \sqrt{\frac{(2J+1)}{32\pi^2 \Delta(\mathfrak{n},K)}} \Big\{ \left[\psi_V + (-i)^K \bar{\psi}_V\right] D_{-KM}^{J*}(\chi,\theta,\varphi) \nonumber \\ & +& (-1)^{J+K+\nu_3}\left[\psi_V + i^K \bar{\psi}_V\right]  D_{KM}^{J*}(\chi,\theta,\varphi)\Big\}~,
\label{eqn:rovibrationalstates_LO_D4h}
\end{eqnarray}
where $\mathbf{Q}$ denotes the normal coordinates in vector form, $\boldsymbol{\Omega}$ the Euler angles and $\Delta(\mathfrak{n},K) \equiv (\delta_{K0}+1)(1+\delta_{\nu_7\nu_8}\delta_{\nu_9\nu_{10}}\delta_{\nu_{11}\nu_{12}})$ a prefactor. 

Besides, the two terms enclosed by the square brackets on the r.h.s. of Eq.~\ref{eqn:rovibrationalstates_LO_D4h} are mapped one another by time reversal, $\mathscr{T}$, and the~\guillemotleft partner\guillemotright~ of the vibrational state in Eq.~(\ref{eqn:vibrationalstatesLO}), $\bar{\psi}_V$, has been introduced as in Sec.~4-2c of Ref.~\cite{BoM75-II},
\begin{eqnarray}
\bar{\psi}_V =& (-1)^{\sum_{i=4}^{12}\nu_i} & \left[\prod_{j=1}^{6}  \Phi_{\nu_j}(Q_j)\right] \Phi_{\nu_8}(Q_7) \Phi_{\nu_7}(Q_8) \nonumber \\ &\times & \Phi_{\nu_{10}}(Q_9) \Phi_{\nu_{9}}(Q_{10}) \Phi_{\nu_{12}}(Q_{11}) \Phi_{\nu_{11}}(Q_{12})~.
\label{eqn:conjugate_vibrationalstatesLO}
\end{eqnarray}
In Eq.~(\ref{eqn:conjugate_vibrationalstatesLO}), the quanta assigned to the 1D harmonic oscillator wavefunctions of the doubly-degenerate modes are pairwise flipped with respect to their original positions in Eq.~(\ref{eqn:vibrationalstatesLO}), \textit{i.e.} $\nu_7 \leftrightarrow \nu_8$, $\nu_{9} \leftrightarrow \nu_{10}$ and $\nu_{11} \leftrightarrow \nu_{12}$. Moreover, if $\nu_{i}=\nu_{i+1}$ in all the doubly-degenerate modes, the states in Eq.~(\ref{eqn:rovibrationalstates_LO_D4h}) reacquire the factorized form between the vibrational and the rotational part, even when $K \neq 0$.

Since $H_{LO}$ is invariant under parity, $\mathscr{P}$, and time reversal, $\mathscr{T}$, the eigenfunctions of the Hamiltonian in Eq.~\ref{eqn:RigRotHamiltonian} transform according to irreducible representations of the two point groups, isomorphic to the cyclic group of order two, $C_2$, in Sch\"onflies notation \cite{Car97}. Due to $\mathscr{P}$ and $\mathscr{T}$ symmetries, the parity and time-reversal operators do not affect the orientation of the body-fixed frame, \textit{i.e.} the Euler angles. Therefore, the parity $\pi$ of the full eigenstate of $H_{LO}$ depends on the one of the vibrational part of the wavefunctions, $\psi_V$.

At LO, the eigefunctions of the Hamiltonian have well-defined transformation properties under the operations of $\mathcal{D}_{4h}$. Consequently, the $\Psi_{RV}^{\mathcal{D}_{4h}}$'s must transform as a 1D representation completely symmetric (symmetric or antisymmetric) under proper (improper) rotations, due to the bosonic nature of the\ce{^4He} clusters. Indeed, the irreducible representations of ${}^{24}\mathrm{Mg}$ fulfilling these properties are the $A_{1g}$ and the $A_{1u}$ \cite{StS24}, associated with even and odd parity states respectively \cite{Ste26}. This constraint permits to determine the allowed angular momentum states associated with a specific normal-mode excitation and a fixed $K$-value, recapitulated in Tab.~5 of Ref.~\cite{Ste26} as well as in Tab.~I of Ref.~\cite{HWD71}. However, at higher orders in the approximation scheme for the Watson Hamiltonian, triaxiality is induced and the dihedral group $\mathcal{D}_{4h}$ no longer represents an exact symmetry of the Hamiltonian \cite{Ste26}.

Besides the energy eigenstates, for the validation of the $\mathcal{D}_{4h}$-symmetric structural ansatz, it is important to focus on the observables related to $\gamma$-transitions and electron scattering on the \ce{^{24}Mg} nucleus. In the G$\alpha$CM framework, the squared transition form factors between two rotational-vibrational eigenstates with definite $J$, $K$, $M$ and $[\mathfrak{n}] $ are defined as
\begin{eqnarray}
|F(\mathbf{q}, &J_i^{\pi_i},& K_i, [\mathfrak{n}]_i \rightarrow J_f^{\pi_f}, K_f, [\mathfrak{n}]_f)|^2 \equiv  \frac{1}{2J_i+1}\sum_{M_i=-J_i}^{J_i}\sum_{M_f=-J_f}^{J_f} \nonumber \\
&\cdot &  |\langle J_f, M_f, |K_f|, [\mathfrak{n}]_f |  F(\mathbf{q})| J_i, M_i, |K_i|, [\mathfrak{n}]_i \rangle |^2~,
\label{eqn:squareFormFactor}
\end{eqnarray}
where $\mathbf{r}_i \equiv (x_i,y_i,z_i)$ are the $\alpha$-particles' positions in the intrinsic frame, depending on the normal coordinates. With reference to Eq.~\eqref{eqn:squareFormFactor}, 
\begin{equation}
F(\mathbf{q}) \equiv \sum_{\lambda=0}^{+\infty}\sum_{\nu=-\lambda}^{\lambda}f_{\lambda \nu}(q,\mathbf{Q}) D_{\nu 0}^{\lambda}(\chi,\theta,\varphi)~,
\label{eqn:FormFactor_LabFrame}
\end{equation}
is the form-factor operator in the laboratory frame, whereas the intrinsic counterpart is given by
\begin{equation}
f_{\lambda\nu}(q,\mathbf{Q}) \equiv \mathrm{i}^{\lambda} \frac{2\pi}{3} \sum_{i=1}^6 j_{\lambda}(q~r_i) Y_{\lambda}^{\nu*}(\hat{\mathbf{r}}_i)
\label{eqn:FormFactor_BodyfixedFrame}
\end{equation}
where $Y_{\lambda}^{\nu*}$ is a spherical harmonic. Approximations to the exact expression of the squared form-factors can be obtained by evaluating the operator in Eq.~\eqref{eqn:FormFactor_LabFrame} at the equilibrium $\alpha$-particles' positions (static limit), $\mathbf{r}_i^e \equiv (x_i^e,y_i^e,z_i^e)$ as in Refs.~\cite{HWD71,BiI21-01}. Specifically, for a $\mathcal{D}_{4h}$-symmetric configuration and transitions between the ground state and states with zero vibrational quanta, one obtains
\begin{eqnarray}
F_{stat}(\mathbf{q},&0^+,& 0, [\mathbf{0}] \rightarrow J_f^{\pi_f}, K_f, [\mathfrak{n}]_f) \equiv  \frac{\sqrt{4\pi}}{6} j_{J_f}(q \beta_1) \mathrm{f}(q) \sum_{\ell=0}^3 Y_{J_f}^{|K_f|}(\mscriptsize{\frac{\pi}{2}, (2\ell+1)\frac{\pi}{4}}) \nonumber  \\ & + & \frac{\sqrt{4\pi}}{3} j_{J_f}(q \beta_2) \mathrm{f}(q)  \sqrt{\frac{2J_f+1}{4\pi}}\left(\frac{1+(-1)^{J_f}}{2}\right)\delta_{K_f 0}~,
\label{eqn:StaticFormFactor_LabFrame}
\end{eqnarray}
where $j_{J_f}$ is a spherical Bessel function and $\mathrm{f}(r)=1$ for a pointlike charge distribution for the \ce{^4He} clusters and $|K_f| = 0, 4, 8 \ldots$. For a spherical Gaussian profile of charge distribution $\mathrm{f}(r) = \exp(-\frac{r^2}{4\alpha_1})$ with $\alpha_1=0.53~\mathrm{fm}^{-2}$, as in Ref.~\cite{BiI21-01}.

Next, one considers the $\gamma$-transitions between the various energy states lying in the rotational bands of ${}^{24}\mathrm{Mg}$. Although of different nature, electric monopole transitions are often included in the set. The reduced electric ($F = E$) or magnetic ($F = M$) multipole transition probability between the initial state $(\Psi_{RV}^{\mathcal{D}_{4h}})_i$ characterized by the parity $\pi_i$, the quantum numbers $J_i$, $M_i$, $K_i$ and $[\mathfrak{n}]_i$ phonons, and the final state $(\Psi_{RV}^{\mathcal{D}_{4h}})_f$ with parity $\pi_f$ and the quantum numbers $J_f$, $M_f$, $K_f$ and $[\mathfrak{n}]_f$, is provided by
\begin{eqnarray}
B(F\lambda, & J_i^{\pi_i}, & |K_i|, [\mathfrak{n}]_i  \rightarrow  J_f^{\pi_f}, |K_f|, [\mathfrak{n}]_f)
  =  \frac{1}{2J_i+1}\sum_{Mi=-J_i}^{J_i}\sum_{Mf=-J_f}^{J_f} \nonumber \\ & \times & \sum_{\mu = -\lambda}^{\lambda}  |\langle J_f, M_f, |K_f|, [\mathfrak{n}]_f |   \Omega_{\lambda\mu}(F)| J_i, M_i, |K_i|, [\mathfrak{n}]_i  \rangle |^2~,
 \label{eqn:reducedEMtransitionprobabibilites}
\end{eqnarray}
where $\Omega_{\lambda\mu}(F)$ is the transition operator in the laboratory frame, connected with the intrinsic counterpart, $\omega_{\lambda\mu}(F)$, by means of a rotation, 
\begin{equation}
\Omega_{\lambda\mu}(F) = \sum_{\nu = - \lambda}^{\lambda} D_{\mu\nu}^{\lambda}(\varphi,\theta,\chi)~\omega_{\lambda\nu}(F) ~,
\end{equation}
where $D_{\mu\nu}^{\lambda}(\varphi,\theta,\chi)= (-1)^{\mu-\nu} D_{\nu\mu}^{\lambda *}(\chi,\theta,\varphi)$ for Wigner D-matrix elements in the active picture \cite{VdW01}. For electric monopole transitions, triggered by the internal conversion mechanism \cite{KGW22}, the operator $\Omega_{00}(E)$ in the laboratory frame is given by 
\begin{equation}
    \Omega_{00}(E) = 2e\sum_{i=1}^{6}\frac{\xi_i^2 + \eta_i^2 + z_i^2}{\langle r^2 \rangle _{g.s.}}~,
\end{equation}
where $\langle r^2 \rangle _{g.s.}$ is the experimental squared charge radius of the $0^+$ ground state of the nucleus \cite{StS24}. 
Finally, one defines the EM moments which characterize a given rotational-vibrational state of parity $\pi$, angular momentum $J$, projection $|K|$ on the intrinsic frame and $[\mathfrak{n}]$ vibrational quanta. Setting the angular-momentum projection to the maximum value on the laboratory frame, $M=J$, the magnetic dipole moment becomes 
\begin{eqnarray}
\mu( J^{\pi}, |K|, [\mathfrak{n}])
  \equiv  \langle J, J, |K|, [\mathfrak{n}] |   \Omega_{10}(M)| J, J, |K|, [\mathfrak{n}]  \rangle ~,
 \label{eqn:magnetic-dipole-moment}
\end{eqnarray}
whereas the electric quadrupole moment gives
\begin{eqnarray}
Q(  J^{\pi},  |K|, [\mathfrak{n}])
  =  \langle J, J, |K|, [\mathfrak{n}] |   \Omega_{20}(E)| J, J, |K|, [\mathfrak{n}]  \rangle ~.
 \label{eqn:electric-quadrupole-moment}
\end{eqnarray}
Higher-order EM moments will be considered only in Ref.~\cite{Ste26}.

\section{Low-energy spectrum and $\gamma$-transitions}\label{sec:spectrum_EM_transitions}

In the present model, the observed $J^{\pi}$ energy states are grouped into rotational bands, characterized by $|K|^{\pi}$, the number of vibrational quanta $[\mathfrak{n}]$ in the vibrational part of the assigned $\Psi_{RV}^{\mathcal{D}_{4h}}$ states and (a direct sum of) irreducible representations of $\mathcal{D}_{4h}$ associated with the corresponding excited normal mode(s). Representatives for all the singly-excited vibrational modes have been detected \cite{StS24,Ste26}, and the composition of the lowest-$|K|$ bands is mostly coherent with the literature \cite{FKA24,GFM78,FGH79,CLW93,KYI12}. The classification of states into bands with definite $|K|$ value remains valid in good approximation also for models which consider the \ce{^{24}Mg} nucleus as triaxial \footnote{T. Otsuka, \emph{A novel overall view of nuclear shapes, rotations and vibrations}, European Nuclear Physics Conference, Caen (2025)}, such as the $\alpha$-particle models with a $\mathcal{D}_{2h}$-symme\-tric equilibrium structure \cite{HWD71,BSS80}.
However, the classification of negative-parity low-energy levels into rotational bands is less unanimous for this nucleus \cite{KEO21} and the composition of singly-excited bands with higher $|K|$ remains quite speculative, as for part of the member states the corresponding parity or the total angular momentum is uncertain or undetermined. Only new experiments that exploit, for instance, the \ce{^{23}Mg}$(n,\gamma)$\ce{^{24}Mg} or \ce{^{12}C}(\ce{^{12}C},\ce{^{12}C}*)\ce{^{12}C}* reactions \cite{VOF26}, might shed light to the angular momentum and spin of the many documented lines lying between $10$ and $13$ MeV, by performing angular correlation and distribution analysis.

As for $^{12}\mathrm{C}$ \cite{SFV16,BiI00,FSV17}, $^{16}\mathrm{O}$ \cite{BiI17,BiI14} and $^{20}\mathrm{Ne}$ \cite{BiI21-01}, not all the observed low-energy spectrum is susceptible to a description in terms of $^{4}\mathrm{He}$ clusters. Besides the experimental lines characterized by total isospin $T=1$ or $2$, incompatible with the bosonic nature with zero isospin of the $\alpha$ particles, the states of lowest $|K|^{\pi}=3^-$ band \cite{GFM78,BGW71}, composed by the $3_1^-$ line at 7.61641(7) MeV, the $(4)_1^-$ at 9.2998(3) MeV and the $(5)_5^-$ at 11.909(2) MeV, have been treated as non $\alpha$-cluster states and excluded from analysis of the spectrum in Ref.~\cite{Ste26}. In addition, the moment of inertia $\mathscr{I}_x$ of the band is rather small and incompatible with the one of the \textit{g.s.} $|K|^{\pi}=0^+$ band. The reproduction itself of the latter $|K|^{\pi}=3^-$ band proved to be rather challenging for cluster models so far \cite{HWD71,KaH79,FHI80,DeB87,DeB89}. 

Concerning the $\gamma$-transition strengths, the available results of the $\mathcal{D}_{4h}$-symmetric G$\alpha$CM at LO deliver results which are capable of capturing the established trends in the observed rotational bands $|K|^{\pi}$. Nonetheless, the measured EM multipole transition probabilities represent only a small portion of the allowed transitions between the states classified into the $|K|^{\pi}=0^+$ unexcited band and the nine singly-excited rotational bands with lowest $|K|$ value. The intraband transition strengths depend strongly on the structure parameters $\beta_1$ and $\beta_2$, whereas for the interband ones the frequencies of the excited vibrational modes, $\hbar\omega_i$, are crucial. The presence of $\mathcal{D}_{4h}$ symmetry imposes additional selection rules, \textit{i.e.} constraints on the $\gamma$-transition modes based on the vanishing integral rule \cite{BuJ04}, and are discussed extensively in Ref.~\cite{StS24}. The measured E1 transition probabilities cannot be explained in the G$\alpha$ CM framework, as they involve protons and neutrons separately. For the sake of brevity, the results for interband transitions at LO are covered in Ref.~\cite{Ste26} and are left aside this proceedings.

\subsection{Unexcited bands}\label{sec:unexcited_A1g_band}

For the bands corresponding to zero vibrational excitation, $\mathfrak{n}=0$, the G$\alpha$CM predicts rotational bands with $|K|=0, 4, 8 \ldots$, all with positive parity \cite{HWD71,StS24}. The one with $|K|^{\pi}=0^+$ coincides with the ground-state band and is made of states with positive total angular momentum. 
In the observed spectrum, its composition is well-established \cite{BaC22} (cf. Tab.~\ref{tab:composition_unexcited_K=0_band}). 

\begin{table}[htb!]
\centering
\begin{tabular}{ccccc}
\toprule
 \multicolumn{2}{c}{$A_{1g}$ (g.s.) \textsc{Band}} & Exper. [MeV] &  \multicolumn{2}{c}{G$\alpha$\textsc{cm}~LO $(\beta_1,\beta_2)$ [e $\mathrm{fm}^2$]} \\
$J^{\pi}$ & Refs. &  Ref. \cite{BaC22} &  $(2.25,3.57)$ fm & $(2.38,3.86)$ fm \\ 
\midrule
$0^+$ & \cite{GFM78,CLW93,KYI12,ChK15,KEO21} & 0.0(0) & 0.0 & 0.0 \\
$2^+$ & \cite{GFM78,CLW93,KYI12,ChK15,KEO21} & 1.368667(5) & 0.882 & 0.762 \\
$4^+$ & \cite{GFM78,CLW93,KYI12,ChK15,KEO21} & 4.122853(12) & 2.940 & 2.543 \\
$6^+$ & \cite{GFM78,CLW93,KYI12,ChK15,KEO21} & 8.1132(10) & 6.174 & 5.340  \\
$8^+$ & New assignment & 11.860(2) & 10.584 & 9.155 \\
\bottomrule
\end{tabular}
\caption{The ground-state $|K|^{\pi}=0^+$ band of \ce{^{24}Mg}, corresponding to zero quanta of vibrational excitation ($A_{1g}$ irrep). The calculated energy states refer to the parameter sets $(\beta_1,\beta_2) \approx (2.247,3.566)$ and $(2.380,3.857)~\mathrm{fm}$. The composition reflects the one in the NNDC/ENSDF database, except for the inclusion of the $8^+$ line at 11.860 MeV. A measured E2 transition strength between the latter state and the $6^+$ line of the band would support the classification. } \label{tab:composition_unexcited_K=0_band}
\end{table}

The measured nuclear properties include the charge radius of the $0_1^+$ state, equal to $\sqrt{\langle r^2 \rangle _{g.s.}} = 3.0570(16)~\mathrm{fm}$ \cite{AnM13}, the electic quadrupole moment of the $2_1^+$ state, equal to $Q = -29.0(30)~e~\mathrm{fm}^2$ \cite{KSG15} or $-16.6(6)~e~\mathrm{fm}^2$ \cite{Fir07} and the magnetic dipole moments of the $2_1^+$ and $4_1^+$ states, equal to $\mu = -1.08(3)~\mu_N$ \cite{KSG15} and $+1.7(12)~\mu_N$ \cite{KSG15} respectively. Consequently, the nucleus has a marked prolate deformation and $\alpha$-cluster character \cite{SSM22}. 

Fitting the structure paramaters of the nucleus, $\beta_1$ and $\beta_2$ on the experimental charge radius and the E2 moment in Ref.~\cite{Fir07}, one obtains the set $(\beta_1,\beta_2) \approx (2.247,3.566)~\mathrm{fm}$, whereas, with the quadrupole moment in Ref.~\cite{KSG15} one obtains $(\beta_1,\beta_2) \approx (1.812,4.031)~\mathrm{fm}$. Since the charge radius of the $\alpha$-particle is equal to $1.6755(28)~\mathrm{fm}$ \cite{AnM13}, the measurement in Ref.~\cite{Fir07} is more compatible with the chosen square bipyramidal $\alpha$-structure, as it minimizes the overlap between the planar \ce{^4He} clusters as for \ce{^{20}Ne} \cite{BiI21-01}. On side of these sets of structure parameters, one might consider to fix $\beta_1$ on the minimum distance to avoid overlap between the planar $\alpha$-particles and adjust $\beta_2$ on the average between the value wihch delivers the best fit to the \ce{^{24}Mg} charge radius and the E2 moment in Ref.~\cite{KSG15}, obtaining $(\beta_1,\beta_2) \approx (2.380,3.857)~\mathrm{fm}$. 

Recent investigations on the ground state based on the AMD approach combined with the GCM \cite{KYI12,ChK20} suggest that the $0_1^+$ state of the nucleus has a Bohr-Mottelson quadrupole deformation \cite{BoM75-II} parameter $\beta=0.49$, together with a mild triaxiality, $\gamma=13^{\circ}$. In the LO G$\alpha $CM, one obtains $\beta \approx 0.57$ (1.12) from the parameter set based on the electric quadrupole moment  in Ref.~\cite{Fir07} (\cite{KSG15}), whereas from the set $(\beta_1,\beta_2) \approx (2.380,3.857)~\mathrm{fm}$ the value $\beta\approx 0.60$ is found. Consequently, the set $(\beta_1,\beta_2) \approx (1.812,4.031)~\mathrm{fm}$ can be left aside, for reasons of compatibility, also with the early calculations in the framework of Nilsson's microscopic rotational model \cite{Hor72}. In both the three sets $\gamma=0^{\circ}$, since two moments of inertia coincide, cf. Eq.~\eqref{eqn:StatInertiaTensor_pl}. Other AMD applications, predict deformation parameters $(\beta,\gamma) \approx (0.48,22^{\circ})$ \cite{ChK15} and $ (0.35,3^{\circ})$ \cite{KEO21}. In all AMD applications, the intrinsic matter density distribution of the $0_1^+$ state indicates a \ce{^{12}C}+\ce{^{12}C} cluster configuration, recalling the one of a deformed square bipyramid, with the planar $\alpha$-clusters staggered in pairs upwards and downwards with respect to the $z=0$ plane. 

\begin{table}[htb!]
\centering
\begin{tabular}{ccccc}
\toprule
\multirow{2}{2.0cm}{\centering{$A_{1g}$ (g.s.) \textsc{Band}}} & \multicolumn{2}{c}{\textsc{Exper.} [e $\mathrm{fm}^2$]} &  \multicolumn{2}{c}{G$\alpha$\textsc{cm}~LO $(\beta_1,\beta_2)$ [e $\mathrm{fm}^2$]}\\
 &   Ref.~\cite{Fir07} &  Ref.~\cite{KSG15} &  $(2.25,3.57)$ fm & $(2.38,3.86)$ fm\\
\midrule
$Q[2^+~(1.369)]$ & $-16.6(6)$  & $-29.0(30)$ & -16.60 & -20.15 \\
$Q[4^+~(4.123)]$ & n.a.  & n.a. & -21.13 & -25.64 \\
$Q[6^+~(8.113)]$ & n.a.  & n.a. & -23.24 & -28.21 \\
$Q[8^+~(11.860)]$ & n.a.  & n.a. & -24.46 & -29.69 \\
\bottomrule
\end{tabular}
\caption{Quadrupole moments of the states belonging to the lowest $|K|^{\pi}=0^+$ band of \ce{^{24}Mg}, corresponding to zero quanta of vibrational excitation ($A_{1g}$ irrep). The calculated E2 moments states refer to the parameter sets $(\beta_1,\beta_2) \approx (2.247,3.566)$ and $(2.380,3.857)~\mathrm{fm}$.} \label{tab:electric-quadrupole-moments_gsBand}
\end{table}

Regarding the moments of inertia, the deviation between the \textit{rigid-rotor} estimate and the counterpart obtained from the interpolation on the experimental energies of the states lying in the $|K|^{\pi}=0^+$ band are, in general, larger than the statistical errors. Similar discrepancies are highlighted in Ref.~\cite{BiI21-01}, where the same approach has been applied to LO for \ce{^{20}Ne}. Specifically, for the ground-state band of \ce{^{24}Mg}, one has a nucleon-mass-specific moment of inertia \cite{BiI21-01} with respect to the x (z) axis of 125.5(78)~fm$^2$ (24.0(131)~fm$^2$). For the same band, with the structure parameters $(\beta_1,\beta_2)$ adjusted to $(2.247,3.566)$ and $(1.812,4.031)$ fm, one obtains 142.1 fm$^2$ (80.8~fm$^2$) and  156.3 fm$^2$ (52.5~fm$^2$) respectively. For the fit of the moment of inertia along the $z$ axis, the sequence headed by the $4^+$ state at 8.43929(5) MeV and followed by a $(6)^{(+)}$ at 12.7333(6) MeV, a $8^+$ state at 17.90(1) MeV and a $(10)^{(+)}$ state at 23.26(1) MeV has been exploited as $|K|^{\pi}=4^+$ sideband \cite{Bou62}.

Making use of the identity in Eq.~\eqref{eqn:electric-quadrupole-moment}, the electric quadrupole moment has been calculated for all the detected states of the lowest $|K|^{\pi}=0^+$ band (cf. Tab.~\ref{tab:electric-quadrupole-moments_gsBand}). Despite the fact that $\beta_2$ in the parameter set $(\beta_1,\beta_2) \approx (2.380,3.857)~\mathrm{fm}$ is obtained from an average between the $\beta_2$ value wihch gives the best fit to the \ce{^{24}Mg} charge radius and the E2 moment in Ref.~\cite{KSG15}, the E2 moment estimate for the $2_1^+$ state is in better agreement with the measurement in Ref.~\cite{Fir07}.
 
\begin{figure}[htb!]
\begin{center}
\begin{tikzpicture} 
    \begin{axis}[
    	xlabel={$q~[\mathrm{fm}^{-1}]$},
    	ylabel={$|F_{\mathrm{ch}}(q; 0_1^+)|^2$},
	height = 7.955 cm,
           width = 11.255 cm,
	legend style ={at={(0.03,0.41)}, anchor=north west, draw=black,fill=white,align=left},
	legend entries ={\textcolor{Black}{\small{ Exp. \cite{MaM89} }}, \textcolor{Black}{\small{Fit Exp. \cite{IsF25}}}, \textcolor{Black}{\small{G$\alpha$CM LO (Dyn.)}},  \textcolor{Black}{\small{Stat. Pointlike}},  \textcolor{Black}{\small{Stat. Gaussian}}},
	legend columns = 1,
   	 xmin=0.0,
	xmax=4.0,
   	ymin=0.0000000001, 
          ymode=log,
	ymax=1.25,
	xtick={0.0, 0.50, 1.00, 1.50, 2.00,  2.50,  3.00, 3.50, 4.00},
	ytick={0.0000000001, 0.000000001, 0.00000001,0.0000001,0.000001,0.00001,0.0001,0.001,0.01,0.1,1.0}]
    \addplot[
    scatter/classes={a={Bittersweet}},
    scatter,
    only marks,
    scatter src=explicit symbolic
    ]
    plot [error bars/.cd, y dir = both, y explicit]
    table[meta=Colour, x=q, y=Exp0EFF, y error=ExpE0FFErr]{
	q	Exp0EFF		ExpE0FFErr	Colour
	0.68764	0.159612000		0.0175615000	a
	0.782022	0.10637000		0.0165545000	a
	0.930337	0.04893900		0.0065016300	a
	1.01798	0.027825600		0.0036598500	a
	1.09888	0.0150053000	0.0020146200	a
	1.23371	0.002908050		0.0005081610	a
	1.30112	0.000684304		0.0001011030	a
	1.37598	0.000259294		0.0000741852	a
	1.42921	0.00032044		0.0000445275	a
	1.50337	0.000480831		0.0000743261	a
	1.61124	0.000649024		0.0000513421	a
	1.65843	0.000788046		0.0000674514	a
	1.70562	0.000802074		0.0001019750	a
	1.77303	0.000544042		0.0000551243	a
	1.92809	0.000448065		0.0000451343	a
	2.13708	0.0000740853	0.00000901524	a
	2.17079	0.000155443		0.0000204763	a
	2.34607	0.0000165345	0.00000209563	a
    };
    \addplot[
 	scatter/classes={a={Bittersweet}},
    scatter,
    no marks,
    draw=Bittersweet,
    scatter src=explicit symbolic
    ]
    plot [error bars/.cd, y dir = both, y explicit]
    table[x=q, y=Best0EFF]{
	q	Best0EFF
	0.23939393939393938	0.8354491640808372	
	0.296969696969697	0.7413901810893306	
	0.34545454545454546	0.6322408675187504	
	0.396969696969697	0.5610600757228686	
	0.4484848484848485	0.47845941051439583	
	0.49696969696969695	0.42459209280008264	
	0.5484848484848485	0.34794978349931255	 
	0.6000000100000000	0.2740123724483524	 
	0.6454545454545455	0.2245509214285554	
	0.696969696969697	0.1768350883776987	
	0.7484848484848485	0.12859968758886134	
	0.7999999999998000	0.10127296283375798	
	0.8454545454545455	0.0707740289396849	
	0.899999978990000	0.04946002400045896	
	0.9454545454545454	0.03191924378134716	
	0.996969696969697	0.01979519563724472 	
	1.0484848484848486	0.01227628614891497 	
	1.0969696969696970	0.006756176841991144 	
	1.1454545454545455	0.0032996039555468707	
	1.200001912000000	0.0014300440772638799 
	1.2515151515151515	0.000433129171958053	
	1.2969696969696969	0.000050454494098247	
	1.3484848484848484	0.000012523119823862	
	1.4000010101000000	0.000136513786520062	
	1.4515151515151514	0.000302689836389009	
	1.496969696969697	0.0004507217350319378	
	1.5484848484848486	0.0005500014479763943	
	1.600001020300000	0.0005955880363182272	
	1.6484848484848484	0.0005955880363180272	
	1.700010000000000	0.0005500014479763943	
	1.7515151515151515	0.000488079575210587	
	1.800002340000000	0.000399977261481785	
	1.8484848484848484	0.0003149842981414334	
	1.900000000001000	0.000238369914620509	
	1.9484848484848485	0.00016658344698965	
	1.999999999999990	0.00011187194314213	
	2.0484848484848484	0.00007219704999362	
	2.100000000010000	0.00004302645750181	
	2.151515151515151	0.00002275508857012	
	2.196969696969697	0.00000986202589030	
	2.251515151515151	0.00000310831637351	
	2.300000000100000	0.00000037678942311	
	2.348484848484849	0.00000010127296284	
	2.400000010000000	0.00000110397056189	
	2.448484848484848	0.00000254724011010	
	2.496969696969697	0.0000039470407834026	
	2.5515151515151513	0.00000481644876953	
	2.6000003012000000	0.00000521565765913	
	2.6484848484848484	0.00000521565765913	
	2.703030303030303	0.00000481644876953	
	2.745454545454545	0.00000427418923732	
	2.796969696969697	0.00000336594758426	
	2.851515151515152	0.00000265070227613	
	2.900000010100100	0.00000192766881125	
	2.951515151515152	0.00000129455873045	
	2.996969696969697	0.00000080283992656	
    };
 \addplot[
 	scatter/classes={a={Cerulean}},
    scatter,
    no marks,
    line width=0.7mm,
    draw=Cerulean,
    scatter src=explicit symbolic
    ]
    plot [error bars/.cd, y dir = both, y explicit]
    table[x=q, y=GaCM0EFF]{
	q	GaCM0EFF	
	0.00	1.0000000	
	0.10	0.965146	
	0.20	0.8671335	
	0.30	0.723749	
	0.40	0.559103	
	0.50	0.397424	
	0.60	0.2576245	
	0.70	0.1502015	
	0.80	0.0771652	
	0.90	0.0338816	
	1.00	0.0129639	
	1.10	0.002413691	
	1.20	0.0001054573 
	1.22	0.00001973258	
	1.24	0.0000005063020	
	1.26	0.00002787761	
	1.28	0.0000857330	
	1.30	0.0001625075
	1.35	0.0003742162	
	1.40	0.0005416421
	1.45	0.000623526	
	1.50	0.0006241677
	1.55	0.0005562421	
	1.60	0.0004570125
	1.65	0.0003404463
	1.70	0.00023737700
	1.75	0.0001477585	
	1.80	0.00008406829
	1.85	0.00004187812	
	1.90	0.00001761331	
	1.95	0.000005616778
	2.00	0.000001056039	
	2.05	0.000000398912
	2.10	0.000000112549	
	2.15	0.000000065856
	2.20	0.0000000491175
	2.25	0.0000002272054	
	2.30	0.0000006184220	
	2.35	0.000002142001
	2.40	0.000004517847
	2.45	0.000007740103	
	2.50	0.00001107460
	2.55	0.00001437793	
	2.60	0.00001675997	
	2.70	0.00001850383	
	2.80	0.00001583803	
	2.90	0.00001062363	
	3.00	0.000005383149	
	3.10	0.000001800305	
	3.20	0.000000220699	
	3.28	0.000000006476887	
	3.30	0.00000004094051	
	3.40	0.00000040510860	
	3.50	0.00000070833410	
	3.60	0.00000073847470	
	3.70	0.00000055883760	
	3.80	0.00000032046380	
	3.90	0.00000013417890	
	4.00	0.00000003432484	
    };
      \addplot[Violet, thin, samples = 100, smooth,domain=0.0001:4.0] {(4.0 * sin(deg(2.38 * \x)) / (6.0 * 2.38 * \x)  + 1.0 * sin(deg(3.81 * \x)) / (3.0 * 3.81 * \x ))^2 };
      \addplot [Hotmagenta, thin, samples = 100, smooth,domain=0.0001:4.0] {(4.0 * exp(- \x^2 / (4.0 * 0.53)) * sin(deg(2.38 * \x)) / (6.0 * 2.38 * \x)  + 1.0 * exp(- \x^2 / (4.0 * 0.53)) * sin(deg(3.81 * \x)) / (3.0 * 3.81 * \x ))^2 };
    \end{axis}
    \end{tikzpicture}
\end{center}
\vspace{-0.2cm}
    \caption{
	Squared charge form factor of the ground state $0^+$ of the \ce{^{24}Mg} nucleus. The measured dataset in Ref.~\cite{MaM89} is superimposed by its fit in Ref.~\cite{IsF25} (thin orange curve). The theoretical results in the static limit in Eq.~\eqref{eqn:StaticFormFactor_LabFrame} for pointlike (thin purple curve) and spherical Gaussian (thin magenta curve) $\alpha$-clusters are superimposed, together with the G$\alpha$CM counterpart at LO with the parameter set $(\beta_1,\beta_2) \approx (2.380,3.808)~\mathrm{fm}$ (thick light blue curve).
    }
    \label{fig:ChargeFormFactor_0+_I}
\end{figure}
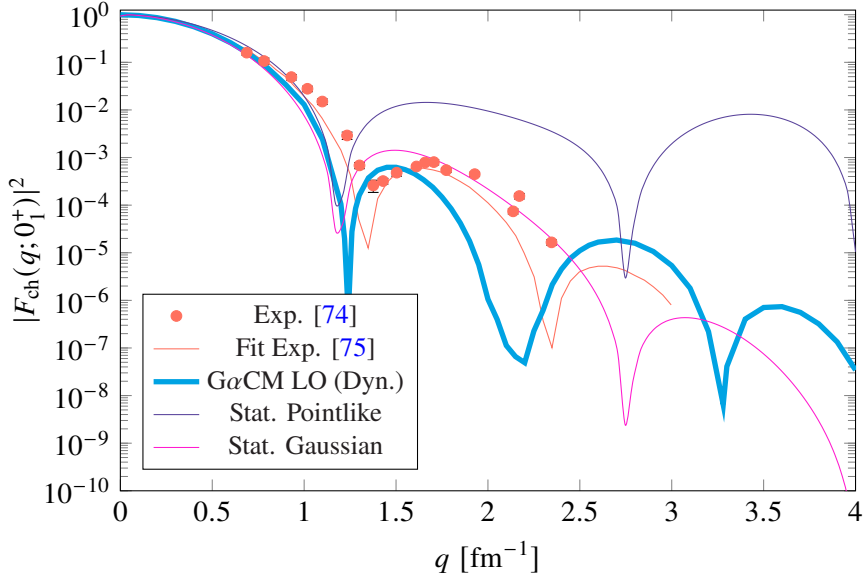

Furthermore, the elastic form factor for the $0_1^+$ state, $F_{\mathrm{ch}}(q;0_1^+)$, has been analyzed, by considering the limit $J_i = J_f = 0$ in the expression in Eq.~\eqref{eqn:FormFactor_LabFrame}, which represents the \textit{dynamic} limit of the one in Eq.~\eqref{eqn:StaticFormFactor_LabFrame}. Taking the squared modulus of the latter, one obtains the curves in Fig.~\ref{fig:ChargeFormFactor_0+_I}, in which the experimental counterpart is taken from Ref.~\cite{MaM89}. The fit of the displayed experimental dataset performed in Ref.~\cite{IsF25}, highlights a quite good agreement of the G$\alpha$CM LO prediction with the observed minima at $q \approx 1.35$ and $2.40~\mathrm{fm}^{-1}$, with deviations of $0.08$-$0.15~\mathrm{fm}^{-1}$. A similar accuracy for the first minimum is reached by the Gaussian static approximation in Eq.~\eqref{eqn:StaticFormFactor_LabFrame}, but for the second minimum the effects of the dynamics, \textit{i.e.} the normal coordinates, are evident. Overall, the two curves deliver a good order-of-magnitude estimate of $|F_{\mathrm{ch}}(q;0_1^+)|^2$. In constrast, the pointlike approximation for the $\alpha$-particles is incapable of capturing the correct order of magnitude of the squared elastic form factor for $q \gtrsim 1.2~\mathrm{fm}$. 

Concerning the transition form factors, the analysis is carried out in Ref.~\cite{Ste26}. At LO in the G$\alpha$CM, an accurate description of $|F(q;0_1^+\rightarrow 2_1^+)|^2$, the observed squared transition form factor to the $2^+$ state of the ground-state band \cite{HHH87}, cannot be achieved with the chosen sets of structure parameters. The position of the first minimum is, in fact, shifted by $\approx 0.5~\mathrm{fm}^{-1}$ towards the origin of the axes. The experimental data \cite{LiY74,MaM89} follow the predicted curve only up to $q \approx 1.0~\mathrm{fm}^{-1}$. For this form factor, the triaxial $\mathcal{D}_{2h}$-symmetric bitetrahedral structure for the $\alpha$-clusters proposed in Ref.~\cite{HWD71} is known to deliver a better description of the position of the minima as well as the global behaviour of the curve. This is also due to the larger versatility of the $\mathcal{D}_{2h}$-symmetric configuration, which depends on three adjustable structure parameters \cite{HWD71}. Preliminary calculations made on a deformed $\mathcal{D}_{4h}$ symmetric equilibrium configuration with planar $\alpha$-clusters staggered pairwise upwards and downwards with respect to the $z=0$ plane by an angle of $\approx 30^{\circ}$ and $(\beta_1,\beta_2)=(2.19,3.09)~\mathrm{fm}$ yield a very accurate description of both the measured $|F_{\mathrm{ch}}(q;0_1^+)|^2$ and $|F(q;0_1^+\rightarrow 2_1^+)|^2$ in the interval $0 \leq q \leq 3~\mathrm{fm}^{-1}$. Such a configuration might mimic the effect of NLO corrections to the LO $\mathcal{D}_{4h}$-symmetric squared form factors, since the rotation-vibration coupling terms introduce a certain degree of triaxiality. At NLO, a new fit of the structure parameters $\beta_1$ and $\beta_2$ is envisaged, in order to reproduce the $0_1^+$ charge radius and the E2 moment of the $2_1^+$ state. Further transition form factors are going to be analysed in Ref.~\cite{Ste26}, thus enabling the comparison with the AMD results in Figs. 5-6 of Ref.~\cite{KEO21}.

\begin{table}[htb!]
\centering
\begin{tabular}{ccccc}
\toprule
\multirow{2}{4.5cm}{\centering{ \textsc{Intraband} $A_{1g}$ (g.s.)}} &  \multicolumn{2}{c}{\textsc{Experimental} \cite{BaC22}}  & \multicolumn{2}{c}{G$\alpha$\textsc{cm}~LO $(\beta_1,\beta_2)$ [$\mathrm{e}^2~\mathrm{fm}^{4}$]}\\
& [W.u.] & [$\mathrm{e}^2~\mathrm{fm}^{4}$] &  $(2.25,3.57)$ fm & $(2.38,3.86)$ fm \\
\midrule
$\mathrm{B}[E2; 0^+~ (0.0) \rightarrow 2^+~(1.369)]$ & $105.5_{-230}^{+240}$ & $433.24_{-946}^{+987}$ & 335.78 & 494.62\\
$\mathrm{B}[E2; 2^+~(1.369) \rightarrow 4^+~(4.123)]$ & $50.0_{-40}^{+47}$ & $205.5_{-167}^{+196}$ & 172.69 & 254.38 \\
$\mathrm{B}[E2; 4^+~(4.123) \rightarrow 6^+~(8.113)]$ & $48.9_{-13}^{+23}$ & $201.1_{-53}^{+95}$ & 152.63 & 224.83 \\
$\mathrm{B}[E2; 6^+~ (8.113) \rightarrow 8^+~(11.860)]$ & $\mathrm{n.a.}$ & $\mathrm{n.a.}$ & 144.64 & 213.07 \\
\bottomrule
\end{tabular}
\caption{Reduced E2 transition probabilities among the states of the ground-state band of \ce{^{24}Mg}, corresponding to zero quanta of vibrational excitation. In the limit $\beta_1 = \beta_2$, the G$\alpha$CM predictions at LO vanish.}\label{tab:EMtransitions_intraband_gsBand_A1g}
\end{table}

Next, the intraband $\gamma$-transitions have been analysed. For processes of elecric quadrupole type, the application of the identity in Eq.~\eqref{eqn:reducedEMtransitionprobabibilites} delivers the results in Tab.~\eqref{tab:EMtransitions_intraband_gsBand_A1g} for the parameter sets $(\beta_1,\beta_2) \approx (2.247,3.566)$ and $(2.380,3.857)~\mathrm{fm}$. Higher-order electric even multipolarities have not been measured yet, as well as M1 or M3 transitions. For the parameter set $(\beta_1,\beta_2) \approx (2.380,3.857)~\mathrm{fm}$, the displayed E2 reduced transition probabilities are in excellent agreement with the experimental counterparts. 
The other structure parameter set ensures at least order-of-magnitude agreement with the measured transition strengths, with an underestimation of 20-25 \% on average. In the AMD application in Ref.~\cite{KYI12}, the reduced E2 transition probabilities of the lowest $|K|=0$ band are reproduced with an accuracy ranging from 4 to 26\%. 

\subsection{First excited $A_{1g}$ band}\label{sec:excited_A1g_I_band}

The first excited $|K|^{\pi}=0^+$ rotational band has been identified as a band of $A_{1g}$ type, corresponding to one quantum of vibrational excitation of energy $\hbar\omega_2 = 6.4322(10)$ MeV, the breathing mode ($\mathfrak{n}_2=1$). In the large-amplitude-vibration limit, the normal mode associated with this band, $\omega_2$, favours the $6\alpha$-cluster decay channel \cite{CRJ23}. The composition of the band in Tab.~\ref{tab:composition_excited_I_K=0+_band} agrees with Refs.~\cite{CLW93,BCH76}, except for the $4^+$ and $6^+$ states, which have been added \cite{StS24,Ste26}. The chosen $2^+$ state does not coincide with the $2^+$ at 7.34860(10) MeV, mentioned in Ref.~\cite{GFM78} as a possible $K^{\pi}=0^+$ state. Instead, the latter has been considered as the bandhead for the $\mathfrak{n}_5=1$ band ($B_{2g}$ type) with $K^{\pi} = 2^+$, in agreement with Ref.~\cite{CLW93}. In Ref.~\cite{KaH79}, both the $2^+$ states at 7.348 and 8.654 MeV are considered as possible members of this band.

\begin{table}[htb!]
\centering
\begin{tabular}{ccccc}
\toprule
 \multicolumn{2}{c}{$A_{1g}$ ($\omega_2$) \textsc{Band}} & Exper. [MeV] &  \multicolumn{2}{c}{G$\alpha$\textsc{cm}~LO $(\beta_1,\beta_2)$ [e $\mathrm{fm}^2$]} \\
$J^{\pi}$ & Refs. &  Ref. \cite{BaC22} &  $(2.25,3.57)$ fm & $(2.38,3.86)$ fm \\ 
\midrule
$0^+$ & \cite{GFM78,CLW93,KYI12,ChK15} & 6.4322(10) & 6.432 & 6.432 \\
$2^+$ & \cite{GFM78,CLW93,KYI12,KEO21,BaC22}  & 8.6545(2) & 7.314 & 7.195\\
$4^+$ & New & 10.66017(17) & 9.372 & 8.975 \\
$6^+$ & New & 15.533(1) & 12.606 & 11.773 \\
\bottomrule
\end{tabular}
\caption{The first excited $|K|^{\pi}=0^+$ band of \ce{^{24}Mg}, corresponding to one quanta of vibrational excitation of frequency $\omega_2$ ($A_{1g}$ irrep). The calculated energy states refer to the parameter sets $(\beta_1,\beta_2) \approx (2.247,3.566)$ and $(2.380,3.857)~\mathrm{fm}$. } \label{tab:composition_excited_I_K=0+_band}
\end{table}

From an interpolation to the energy eigenvalues, one obtains a mass-specific moment of inertia about the intrinsic x-axis of $94.5(78)~\mathrm{fm}^2$, whereas the one about the $z$-axis is unavailable as the $|K|^{\pi} = 4^+$ or $8^+$ sidebands have not been identified yet. The moment of inertia might testify a uniform decrease in the nuclear volume or in the axial deformation, as the one predictable for a $\omega_1$ mode, but the statistical errors are quite large. The alternative choice in the $2^+$ state (cf. Ref.~\cite{GFM78}), instead, would lead to an increased moment of inertia along the x axis from the fit. Nonetheless, in the G$\alpha$CM at LO, the moments of inertia does not depend on the excited vibrational quanta, therefore they are the same as for the ground-state band. Overall, NLO corrections could help establishing the nature of this excited band, since its classification as a $\omega_1$ mode ($A_{1g}$ irrep) can not be excluded at this stage.

\begin{table}[htb!]
\centering
\begin{tabular}{ccccc}
\toprule
\multirow{2}{4.5cm}{\centering{ \textsc{Intraband} $A_{1g}$ ($\omega_2$)}} &  \multicolumn{2}{c}{\textsc{Experimental} \cite{BaC22}}  & \multicolumn{2}{c}{G$\alpha$\textsc{cm}~LO $(\beta_1,\beta_2)$ [$\mathrm{e}^2~\mathrm{fm}^{4}$]}\\
& [W.u.] & [$\mathrm{e}^2~\mathrm{fm}^{4}$] &  $(2.25,3.57)$ fm & $(2.38,3.86)$ fm \\
\midrule
$\mathrm{B}[E2; 0^+~ (6.432) \rightarrow 2^+~(8.655)]$ & $45.0_{-110}^{+160}$ & $185.1_{-452}^{+658}$ & 333.43 & 491.78 \\
$\mathrm{B}[E2; 2^+~(8.655) \rightarrow 4^+~(10.660)]$ & $\mathrm{n.a.}$ & $\mathrm{n.a.}$ & 171.48 & 252.92 \\
$\mathrm{B}[E2; 4^+~(10.660) \rightarrow 6^+~(15.533)]$ & $\mathrm{n.a.}$ & $\mathrm{n.a.}$ & 151.56 & 223.57 \\
\bottomrule
\end{tabular}
\caption{Reduced E2 transition probabilities among the states of the lowest excited $|K|^{\pi}=0^+$ band of \ce{^{24}Mg}, corresponding to one quantum of vibrational excitation of frequency $\omega_2$. }\label{tab:EMtransitions_intraband_I_Band_A1g}
\end{table}

For this band, only one measured intraband reduced transition probability is available, the E2 between the bandhead and the $2^+$ state. From comparison with the experimental value, one infers that the transition rate is overestimated and only order-of-magnitude agreement is found (cf. Tab.~\ref{tab:EMtransitions_intraband_I_Band_A1g}). This could support the fact that the cluster structure is more compact for this band or the quadrupole deformation decreases with respect to the ground-state band. However, the statistical error associated with the measured reduced E2 transition probability is sizable ($\approx 35 \%$). Any further intraband E2 transition strength could corroborate the new assignments, which remain speculative at present.

\subsection{Second excited $A_{1g}$ band}\label{sec:excited_A1g_II_band}

The second excited $|K|^{\pi}=0^+$ rotational band has been classified as a band of $A_{1g}$ type, corresponding to one quantum of vibrational excitation of energy $\hbar\omega_1 = 9.30539(24)$ MeV, the asymmetric stretching mode ($\mathfrak{n}_1=1$). In the large-amplitude-vibration limit, the normal mode associated with this band, $\omega_1$, enhances the $2\alpha+{}^{16}\mathrm{O}$ cluster configuration and decay channel \cite{CRJ23}. As shown in Tab.~\ref{tab:composition_excited_II_K=0+_band}, the bandhead is provided by the $0_3^+$ state at 9.30539(24) MeV, and the other member states partially follow the third $K^{\pi}=0^+$ band in Refs.~\cite{CLW93,CRJ23}. The chosen $2^+$ level does not coincide with the one chosen in Ref.~\cite{CLW93}. The band has been analyzed also in the OCM framework by Ref.\cite{KaH79}, wherein the observed energies are reproduced with average deviations of $\approx 2$ MeV.

\begin{table}[htb!]
\centering
\begin{tabular}{ccccc}
\toprule
 \multicolumn{2}{c}{$A_{1g}$ ($\omega_1$) \textsc{Band}} & Exper. [MeV] &  \multicolumn{2}{c}{G$\alpha$\textsc{cm}~LO $(\beta_1,\beta_2)$ [e $\mathrm{fm}^2$]} \\
$J^{\pi}$ & Refs. &  Ref. \cite{BaC22} &  $(2.19,3.54)$ fm & $(2.38,3.81)$ fm \\ 
\midrule
$0^+$ & \cite{CLW93,ChK15} & 9.30539(24) & 9.305 & 9.305  \\
$2^+$ & New  & 10.6598(2) & 10.210 & 10.082 \\
$4^+$ & \cite{CLW93} & 11.6982(2) & 12.320 & 11.895  \\
$6^+$ & \cite{KYI12,CRJ23} & 16.070(20) & 15.636 & 14.743   \\
\bottomrule
\end{tabular}
\caption{The first excited $|K|^{\pi}=0^+$ band of \ce{^{24}Mg}, corresponding to one quanta of vibrational excitation of frequency $\omega_2$ ($A_{1g}$ irrep). The calculated energy states refer to the parameter sets $(\beta_1,\beta_2) \approx (2.247,3.566)$ and $(2.380,3.857)~\mathrm{fm}$. } \label{tab:composition_excited_II_K=0+_band}
\end{table}

The interpolation to the energy eigenvalues yields a mass-specific moment of inertia about the intrinsic x-axis of $146.3(159)~\mathrm{fm}^2$. As for the first excited $A_{1g}$ band, the moment of inertia about the $z$-axis is unavailable, as $|K|^{\pi} = 4^+$ or $8^+$ sidebands have not been constructed yet. The fitted moment of inertia could witness a uniform increase in the nuclear volume or in the quadrupole deformation, as the one expectable from the breathing mode ($\omega_2$). The latter, in molecules usually appears at lower energy than asymmetric stretching modes \cite{BBR15}. 

Experimental reduced E2 transition probabilities between the member states are absent and G$\alpha$CM predictions at LO have not been calculated yet. Any measured intraband transition strength would help asssessing the composition of the band.

\subsection{First excited $A_{2u}$ band}\label{sec:excited_A2u_I_band}

The first excited $|K|^{\pi}=0^-$ rotational band, has been classified as a band of $A_{2u}$ type, corresponding to one quantum of vibrational excitation of energy $\hbar\omega_3 = 7.3722(10)$ MeV, the symmetric wagging mode ($\mathfrak{n}_3=1$). Due to $\mathcal{D}_{4h}$-symmetry prescpritions, this band is not expected to contain a $J^{\pi}=0^-$ state \cite{HWD71,StS24}. Consequently, the bandhead is the $1^-$ state at 7.5553(10)~MeV, followed by odd-angular momentum states (cf. Tab.~\ref{tab:composition_excited_I_K=0-_band}), as in Refs.~\cite{GFM78,FGH79,CLW93,KYI12,KEO21}. In the large-amplitude-vibration limit, the normal mode associated with this band, $\omega_3$, boosts the $\alpha+{}^{20}\mathrm{Ne}$ cluster configuration and decay channel. Conversely, in the AMD framework, it has been found that the lowest $K^{\pi}=0^-$ band is dominated by the \ce{^{8}Be}+\ce{^{16}O} cluster configuration \cite{ChK20}. 

\begin{table}[htb!]
\centering
\begin{tabular}{ccccc}
\toprule
 \multicolumn{2}{c}{$A_{2u}$ ($\omega_3$) \textsc{Band}} & Exper. [MeV] &  \multicolumn{2}{c}{G$\alpha$\textsc{cm}~LO $(\beta_1,\beta_2)$ [e $\mathrm{fm}^2$]} \\
$J^{\pi}$ & Refs. &  Ref. \cite{BaC22} &  $(2.25,3.57)$ fm & $(2.38,3.86)$ fm \\ 
\midrule
$1^-$ & \cite{GFM78,CLW93,ChK20,KEO21} & 7.5553(10) & 7.666 & 7.627 \\
$3^-$ & \cite{GFM78,CLW93,KEO21} & 8.3581(3) & 8.254 & 8.898 \\
$5^-$ & \cite{FGH79,CLW93,KEO21} & 10.02797(9) & 11.782 & 11.187 \\
$7^-$ & \cite{FGH79,CLW93,KYI12} & 12.443(3) & 15.604 & 14.493  \\
\bottomrule
\end{tabular}
\caption{The first excited $|K|^{\pi}=0^-$ band of \ce{^{24}Mg}, corresponding to one quanta of vibrational excitation of frequency $\omega_3$ ($A_{2u}$ irrep). The calculated energy states refer to the parameter sets $(\beta_1,\beta_2) \approx (2.247,3.566)$ and $(2.380,3.857)~\mathrm{fm}$. } \label{tab:composition_excited_I_K=0-_band}
\end{table}

Concerning the fitted mass-specific moment of inertia about the intrinsic x (z), the value $226.5(52)~\mathrm{fm}^2$ ($66.1(14)~\mathrm{fm}^2$) is found. They suggest a significant growth in the nuclear radius with respect to the ground-state band. As a consequence, the G$\alpha$CM at LO tends to overestimate the energies of the member states in Tab.~\ref{tab:composition_excited_I_K=0-_band}, but the overall accuracy is comparable with the OCM \cite{KaH79}. The AMD calculations in Ref.~\cite{KYI12} predicts for the $1_1^-$ bandhead a larger deformation than for the ground state, $\beta=0.60$, with an axially-symmetric structure, $\gamma \approx 0^{\circ}$. The same trend is captured by the AMD calculations in Ref.~\cite{KEO21}, although with a different $(\beta,\gamma)$ parameter set.

Candidates for the possible $|K|^{\pi}=4^-$ sideband have been identified. The bandhead of the latter is the $J^{\pi} = (4)^-$ state at $12.659(1)~\mathrm{MeV}$ \cite{AbD91}, followed by the $5^-$ at $13.771(3)~\mathrm{MeV}$, the $7^-$ at $15.750(15)~\mathrm{MeV}$ and the $9^-$ at $19.21(4)~\mathrm{MeV}$. 

Among the members of the $|K|^{\pi}=0^-$ band, the only measured transition strength is of E2 type and involves the $3^-$ and $5^-$ states (cf. Tab.~\ref{tab:EMtransitions_intraband_Band_A2u}). The ensuing reducted E2 transition probability in the G$\alpha$CM at LO with the parameter set $(\beta_1,\beta_2) \approx (2.380,3.857)~\mathrm{fm}$ is in excellent agreement with the experimental result, whereas the one with $(\beta_1,\beta_2) \approx (2.247,3.566)$ is underestimated by $\approx 25\%$, thus favouring deformation along the intrinsic $z$ axis. The measurement of the E2 transition strength between the $1^-$ and the $3^-$ would be of interest, together with the one between the $5^-$ state and the $7^-$ at 12.443 MeV, whose predictions in the G$\alpha$CM at LO are reported in Tab.~\ref{tab:EMtransitions_intraband_Band_A2u}.

\begin{table}[htb!]
\centering
\begin{tabular}{ccccc}
\toprule
\multirow{2}{4.5cm}{\centering{ \textsc{Intraband} $A_{2u}$ ($\omega_3$)}} &  \multicolumn{2}{c}{\textsc{Experimental} \cite{BaC22}} & \multicolumn{2}{c}{G$\alpha$\textsc{cm}~LO $(\beta_1,\beta_2)$ [$\mathrm{e}^2~\mathrm{fm}^{4}$]}\\
& [W.u.] & [$\mathrm{e}^2~\mathrm{fm}^{4}$] &  $(2.25,3.57)$ fm & $(2.38,3.86)$ fm  \\
\midrule
$\mathrm{B}[E2; 1^-~(7.555) \rightarrow 3^-~(8.358)]$ & n.a. & n.a. & 196.58 & 290.84 \\
$\mathrm{B}[E2; 3^-~(8.358) \rightarrow 5^-~(10.028)]$ & $50.29_{-126}^{+220}$ & $206.8_{-518}^{+518}$ & 156.02 & 230.83 \\
$\mathrm{B}[E2; 5^-~(10.028) \rightarrow 7^-~(12.443)]$ & n.a. & n.a. & 144.34 & 213.55 \\
\bottomrule
\end{tabular}
\caption{Reduced E2 transition probabilities among the states of the lowest excited $|K|^{\pi}=0^-$ band of \ce{^{24}Mg}, corresponding to one quantum of vibrational excitation of frequency $\omega_3$. }\label{tab:EMtransitions_intraband_Band_A2u}
\end{table}

\subsection{First excited $B_{1g}$ band}\label{sec:excited_B1g_I_band}

As in Ref.~\cite{Bou62}, the first excited $|K|^{\pi}=2^+$ rotational band has been identified as a band of $B_{1g}$ type, corresponding to one quantum of vibrational excitation of energy $\hbar\omega_4 = 2.997(29)$ MeV, an asymmetric stretching mode ($\mathfrak{n}_4=1$). The composition of the band, headed by the $J^{\pi} = 2^+$ state at 4.23835(4) MeV, has firm roots in the literature \cite{Bou62,HaD66,GFM78,CLW93,KYI12} (cf. Tab.~\ref{tab:composition_excited_I_K=2+_band}), although higher-spin members are not included by all the references.

\begin{table}[htb!]
\centering
\begin{tabular}{ccccc}
\toprule
 \multicolumn{2}{c}{$B_{1g}$ ($\omega_4$) \textsc{Band}} & Exper. [MeV] &  \multicolumn{2}{c}{G$\alpha$\textsc{cm}~LO $(\beta_1,\beta_2)$ [e $\mathrm{fm}^2$]} \\
$J^{\pi}$ & Refs. &  Ref. \cite{BaC22} & $(2.25,3.57)$ fm & $(2.38,3.86)$ fm \\ 
\midrule
$2^+$ & \cite{GFM78,CLW93,KYI12,KEO21,BaC22} & 4.23835(4) & 4.326 & 4.174 \\
$3^+$ & \cite{GFM78,CLW93,KYI12,KEO21,BaC22}  & 5.23516(5) & 5.208 & 4.937\\
$4^+$ & \cite{GFM78,CLW93,KYI12,KEO21,BaC22} & 6.01034(5) & 6.384 & 5.954 \\
$5^+$ & \cite{KYI12} & 7.8124(5) & 7.854 & 7.226 \\
$6^+$ & \cite{GFM78,CLW93,KYI12,KEO21,BaC22} & 9.5276(7)  & 9.618 & 8.751 \\
$7^+$ & \cite{GFM78,CLW93,KYI12} & 12.340(15) & 11.676 & 10.531 \\
$8^+$ & \cite{GFM78,CLW93,KYI12} & 14.150(4) & 14.028 & 12.566 \\
$10^+$ & \cite{CLW93} & 19.110(15) & 19.614 & 17.398 \\
\bottomrule
\end{tabular}
\caption{The first $|K|^{\pi}=2^+$ band of \ce{^{24}Mg}, corresponding to one quanta of vibrational excitation of frequency $\omega_4$ ($B_{1g}$ irrep). The calculated energy states refer to the parameter sets $(\beta_1,\beta_2) \approx (2.247,3.566)$ and $(2.380,3.857)~\mathrm{fm}$. The structure reflects the one in the NNDC/ENSDF database, except for the $5^+$, $7^+$, $8^+$ and $10^+$ states. } \label{tab:composition_excited_I_K=2+_band}
\end{table}

The interpolation of the experimental energies delivers a mass-specific moment of inertia about the intrinsic $x$ ($z$) axis of $140.3(15)~\mathrm{fm}^2$ ($87.7(24)~\mathrm{fm}^2$), highlighting a slightly increased nuclear radius, but with reduced deformation than in the ground-state band. The AMD calculation of the energy-density surface in Ref.~\cite{KYI12} for the $2_2^+$ bandhead returns a deformation $\beta \approx 0.44$, smaller than for the ground-state band, in combination with an increased triaixality, $\gamma \approx 14^{\circ}$. The trend is observed also in other states of the same band \cite{KFT12,KiC14,ChK20}, and $|K|$-mixing is often taken into account, also by early calculations \cite{Hor72}. In contrast with the $\mathcal{D}_{4h}$-symmetric G$\alpha$CM, triaxial models such as the ones in Refs.~\cite{HWD71,BSS80} label the lowest $|K|^{\pi} = 2^+$ as a sideband of the ground state band, with zero quanta of collective vibrational excitation. 

According to Ref.~\cite{KEO21} , this $|K|^{\pi}=2^+$ band  is dominated by a $3\alpha$+\ce{^{12}C} cluster structure, although the same series of states have been previously analysed in 2$\alpha$+\ce{^{16}O} configuration \cite{KaH79}.  In addition, candidates for the $|K|^{\pi}=6^+$ sideband have been identified. This band starts with the $J^{\pi} = (6)^+$ state at 12.342(3) MeV and is possibly followed by the $8^+$ at 16.564(10) MeV, the $10^+$ at 22.79 MeV, the $12^+$ at 29.3(1) MeV and the $16^+$ at 46.4(1) MeV, but in absence of transitions the composition remains speculative.

\begin{table}[htb!]
\centering
\begin{tabular}{ccccc}
\toprule
\multirow{2}{4.5cm}{\centering{ \textsc{Intraband} $B_{1g}$ $(\omega_4)$}} &  \multicolumn{2}{c}{\textsc{Experimental} \cite{BaC22}} & \multicolumn{2}{c}{G$\alpha$\textsc{cm}~LO $(\beta_1,\beta_2)$ [$\mathrm{e}^2~\mathrm{fm}^{4}$]}\\
& [W.u.] & [$\mathrm{e}^2~\mathrm{fm}^{4}$] &  $(2.25,3.57)$ fm & $(2.38,3.86)$ fm \\
\midrule
$\mathrm{B}[E2; 2^+~(4.123) \rightarrow 3^+~(5.235)]$ & $\mathrm{n.a.}$ & $\mathrm{n.a.}$ & 170.36 & 250.30 \\
$\mathrm{B}[E2; 2^+~ (4.238) \rightarrow 4^+~(6.010)]$ & $26.82_{-216}^{+216}$ & $110.30_{-888}^{+888}$ & 73.01 & 107.27\\
$\mathrm{B}[E2; 4^+~(6.010) \rightarrow 6^+~(9.528)]$ & $36.1_{-130}^{+317}$ & $148.5_{-535}^{+1307}$ &  115.64 & 169.90 \\
$\mathrm{B}[E2; 5^+~ (7.812) \rightarrow 7^+~(12.340)]$ & $\mathrm{n.a.}$ & $\mathrm{n.a.}$ & 122.53 & 180.04 \\
$\mathrm{B}[E2; 6^+~ (9.527) \rightarrow 8^+~(14.150)]$ & $\mathrm{n.a.}$ & $\mathrm{n.a.}$ & 126.36 & 185.67 \\
$\mathrm{B}[E2; 8^+~ (14.150) \rightarrow 10^+~(19.110)]$ & $\mathrm{n.a.}$ & $\mathrm{n.a.}$ & 129.96 & 190.94 \\
\bottomrule
\end{tabular}
\caption{Reduced E2 transition probabilities among the states of the first $|K|^{\pi}=2^+$ band of \ce{^{24}Mg}, corresponding to one quantum of vibrational excitation of frequency $\omega_4$. }\label{tab:EMtransitions_intraband_Band_B1g}
\end{table}

At present, two experimental reduced E2 transition probabilities between the $|K|^{\pi}=2^+$ member states are available, summarized in Tab.~\eqref{tab:EMtransitions_intraband_Band_B1g}. Analogously to the ground-state band, the G$\alpha$CM LO calculations perform well, with the sample with $(\beta_1,\beta_2) \approx (2.380,3.857)~\mathrm{fm}$ reproducing the measured values with deviations from $2$ to $11\%$. The other set of parameters underestimates the transitions by $21-33\%$. As an outlook, any measured E2 or E4 transition between the $5^+$ line at 7.812 MeV and the other states \cite{GFM78,CLW93,KYI12}, would confirm the classification of this state. The same observation holds for the $7^+$,  $8^+$ and $10^+$ members states, which are enumerated only seldom in literature \cite{GFM78,CLW93,KYI12} as part of the lowest $|K|^{\pi}=2^+$ band (cf. Tab.~\ref{tab:composition_excited_I_K=2+_band}).

\subsection{First excited $B_{2g}$ band}\label{sec:excited_B2g_I_band}

The $J^{\pi} = 2^+$ state at 7.34860(10) MeV has been identified as the head of the second excited $|K|^{\pi}=2^+$ band of $B_{2g}$ type, corresponding to one quantum of vibrational excitation of energy $\hbar\omega_5 = 6.141(55)$ MeV, the symmetric scissoring mode ($\mathfrak{n}_5=1$). The lowest state of this band is classified as a $K=0$ state by Refs. \cite{GFM78,FGH79}, but not by Ref. \cite{CLW93}. The latter $2^+$ state is followed by the $3^+$ state at 9.533 MeV, a $4^+$ state at 10.820 MeV and a $6^+$ state at 14.079 MeV, in agreement with Ref.~\cite{CLW93}. The $(8)^+$ state at 18.16 MeV has been added to the band (cf. Tab.~\ref{tab:composition_excited_II_K=2+_band}).

\begin{table}[htb!]
\centering
\begin{tabular}{ccccc}
\toprule
 \multicolumn{2}{c}{$B_{2g}$ ($\omega_5$) \textsc{Band}} & Exper. [MeV] &  \multicolumn{2}{c}{G$\alpha$\textsc{cm}~LO $(\beta_1,\beta_2)$ [e $\mathrm{fm}^2$]} \\
$J^{\pi}$ & Refs. &  Ref. \cite{BaC22} &  $(2.25,3.57)$ fm & $(2.38,3.86)$ fm \\ 
\midrule
$2^+$ & \cite{CLW93} & 7.34860(10) & 7.533 & 7.317\\
$3^+$ & \cite{CLW93}  & 9.5327(2) & 8.351 & 8.080\\
$4^+$ & \cite{CLW93} & 10.820(8) & 9.527 & 9.097 \\
$6^+$ & \cite{FHS78,CLW93} & 14.079(4) & 12.761&11.895 \\
$(8)^+$ & New & 18.16(1) &17.171 & 15.709 \\
\bottomrule
\end{tabular}
\caption{The second $|K|^{\pi}=2^+$ band of \ce{^{24}Mg}, corresponding to one quanta of vibrational excitation of frequency $\omega_5$ ($B_{2g}$ irrep). The calculated energy states refer to the parameter sets $(\beta_1,\beta_2) \approx (2.247,3.566)$ and $(2.380,3.857)~\mathrm{fm}$. } \label{tab:composition_excited_II_K=2+_band}
\end{table}

Although the level scheme in Ref.~\cite{Bou62} has been adopted, the irreducible representation and the type of excited quanta associated with the present band might be interchanged with the ones of the lowest $B_{1g}$ band. Only the measurement of E2 transition strengths between the states of the second excited $|K|^{\pi}=2^+$ band can resolve the ambiguity on the band type, as well as the one on the bandhead and the $(8)^+$ state. 

The regression over the experimental energies returns a mass-specific moment of inertia about the intrinsic $x$ ($z$) axis of $132.9(58)~\mathrm{fm}^2$ ($92.6(44)~\mathrm{fm}^2$), displaying a slightly increased nuclear radius, but with reduced deformation than in the ground-state band. As for the $B_{1g}$ band, the $|K|^{\pi}=6^+$ sideband has been identified. The latter is headed by the $6^+$ state at 16.203(10) MeV, followed by the $(7)^{(+)}$ state at 18.097 MeV, the $8^+$ at 20.260(15) MeV and the $10^{(+)}$ at 26.80(20) MeV.

Despite the lack of experimental data, some intraband reduced E2 transition probabilities have been calculated in the G$\alpha$CM at LO (cf. Tab.~\ref{tab:EMtransitions_intraband_Band_B2g}). As it can be inferred by comparison with Tab.~\ref{tab:EMtransitions_intraband_Band_B2g}, the E2 transitions are suppressed by $2-5\%$ with respect to the ones for the $B_{1g}$ $|K|^{\pi}=2^+$ band, hence very precise measurements are required in order to ascertain the vibrational mode of this band.

\begin{table}[htb!]
\centering
\begin{tabular}{ccc}
\toprule
\multirow{2}{4.5cm}{\centering{ \textsc{Intraband} $B_{2g}$ ($\omega_5$)}} & \multicolumn{2}{c}{G$\alpha$\textsc{cm}~LO $(\beta_1,\beta_2)$ [$\mathrm{e}^2~\mathrm{fm}^{4}$]}\\
&  $(2.19,3.54)$ fm & $(2.38,3.81)$ fm \\
\midrule
$\mathrm{B}[E2; 2^+~(7.349) \rightarrow 3^+~(9.533)]$ &  172.96 & 253.46 \\
$\mathrm{B}[E2; 2^+~(7.349) \rightarrow 4^+~(10.820)]$ & 74.13 & 108.63 \\
$\mathrm{B}[E2; 3^+~(9.533) \rightarrow 4^+~(10.820)]$ & 118.60 & 173.80 \\
$\mathrm{B}[E2; 4^+~(10.820) \rightarrow 6^+~(14.079)]$ & 117.41 & 172.46 \\
$\mathrm{B}[E2; 6^+~(14.079) \rightarrow 8^+~(18.16)]$ & 128.30 & 188.01 \\
\bottomrule
\end{tabular}
\caption{Reduced E2 transition probabilities among the states of the second $|K|^{\pi}=2^+$ band of \ce{^{24}Mg}, corresponding to one quantum of vibrational excitation of frequency $\omega_5$. }\label{tab:EMtransitions_intraband_Band_B2g}
\end{table}

\subsection{First excited $B_{2u}$ band}\label{sec:excited_B2u_I_band}

The $J^{\pi} = 2^-$ state at 12.2593(5) MeV has been identified as the head of the first excited $|K|^{\pi}=2^-$ band of $B_{2u}$ type, corresponding to one quantum of vibrational excitation of energy $\hbar\omega_6 = 10.721(71)$ MeV. This represents the highest-energy phonon and corresponds to a symmetric twisting mode ($\mathfrak{n}_6=1$).  Although almost unprecedented in the literature, the assignment of this sequence of energy levels, starting with the $2^-$ state at $12.2593(5)$ MeV agrees well with the predictions of antisymmetrized molecular dynamics (AMD) (cf. Fig. 6 of Ref.~\cite{KYI12}). Conversely, the composition disagrees with Ref.~\cite{CLW93}, which considers the same $2^-$ excitation as the member of a $K^{\pi}=1^-$ band, headed by the $1^-$ state at $11.8628(13)$ MeV. 

\begin{table}[htb!]
\centering
\begin{tabular}{ccccc}
\toprule
 \multicolumn{2}{c}{$B_{2u}$ ($\omega_6$) \textsc{Band}} & Exper. [MeV] &  \multicolumn{2}{c}{G$\alpha$\textsc{cm}~LO $(\beta_1,\beta_2)$ [e $\mathrm{fm}^2$]} \\
$J^{\pi}$ & Refs. &  Ref. \cite{BaC22} &  $(2.25,3.57)$ fm & $(2.38,3.86)$ fm \\ 
\midrule
$2^-$ & \cite{CLW93} (?) & 12.2593(5) & 12.049 & 11.897 \\
$3^-$ & \cite{KYI12}   & 14.019(4) & 12.931 & 12.660\\
$5^-$ & \cite{KYI12} & 16.874(6) & 15.577 & 14.949 \\
$7^-$ & \cite{KYI12}  & 19.990(15) & 19.399 & 18.255 \\
$9^-$ & \cite{KYI12} & 25.8(2) & 24.397 & 22.578 \\
\bottomrule
\end{tabular}
\caption{The first $|K|^{\pi}=2^-$ band of \ce{^{24}Mg}, corresponding to one quanta of vibrational excitation of frequency $\omega_6$ ($B_{2u}$ irrep). The calculated energy states refer to the parameter sets $(\beta_1,\beta_2) \approx (2.247,3.566)$ and $(2.380,3.857)~\mathrm{fm}$.  } \label{tab:composition_excited_I_K=2-_band}
\end{table}

Furthermore, Ref.~\cite{CLW93} defines also a $K^{\pi}=0^-$ band, with the $0_1^-$ state at $12.385(1)~\mathrm{MeV}$ and the $2^-$ state at $12.6700(5)$ MeV. However, the only measured $0^-$ state cannot belong to a singly-excited rotational band. Indeed, $J^{\pi}=0^-$ states must have at least two excitation quanta. Specifically, considered the fact that $\hbar\omega_7 + \hbar\omega_9 = 12.29(3)~\mathrm{MeV}$ \cite{Ste26} and that $n_7$ and $n_9$ = 1 states transform as the $E_g \otimes E_u= A_{1u} \oplus A_{2u} \oplus B_{1u} \oplus B_{2u}$ representation, the measured level could constitute the lowest level of a $K^{\pi}=0^-$ band with $J^{\pi}=$ $0^-$, $1^-$, $2^-$, $3^-$, $\ldots$ states as members. 

The interpolation of the experimental energies gives a mass-specific moment of inertia about the intrinsic $x$ ($z$) axis of $133.5(69)~\mathrm{fm}^2$ ($67.5(32)~\mathrm{fm}^2$), brings out a slightly increased nuclear radius, accompanied by a reduced deformation in comparison with the ground-state band. Consequently, the energy eigenvalues are reproduced quite precisely at LO (cf. Tab.~\ref{tab:composition_excited_I_K=2-_band}), with discrepancies from 5 to 12 \%.  The sample with $(\beta_1,\beta_2) \approx (2.189,3.541)$ delivers the most accurate estimates. 

Additionally, the $|K|^{\pi}=6^-$ sideband has been tentatively identified and exploited for the estimation of the moment of inertia about the intrinsic $z$ axis. The latter band is headed by the $(6)^{(-)}$ state at 22.93(3) MeV, followed by the $(8)^{(-)}$ state at 27.4(1) MeV, the $(9)^{(-)}$ at 31.8(1) MeV and the $(10)^{(-)}$ at 33.1(1) MeV. Nonetheless, the angular momentum assignments of the possible member states are still uncertain and the composition of this sideband remains tentative.

\begin{table}[htb!]
\centering
\begin{tabular}{ccc}
\toprule
\multirow{2}{4.5cm}{\centering{ \textsc{Intraband} $B_{2u}$ ($\omega_6$)}} & \multicolumn{2}{c}{G$\alpha$\textsc{cm}~LO $(\beta_1,\beta_2)$ [$\mathrm{e}^2~\mathrm{fm}^{4}$]}\\
&  $(2.25,3.57)$ fm & $(2.38,3.86)$ fm \\
\midrule
$\mathrm{B}[E2; 2^-~(12.259) \rightarrow 3^-~(14.019)]$ &  170.36 & 227.61 \\
$\mathrm{B}[E2; 3^-~(14.019) \rightarrow 5^-~(16.874)]$ & 73.01 & 97.55 \\
$\mathrm{B}[E2; 5^-~(16.874) \rightarrow 7^-~(19.990)]$ & 116.82 & 156.08 \\
$\mathrm{B}[E2; 7^-~(19.990) \rightarrow 9^-~(25.800)]$ & 124.62 & 184.13 \\
\bottomrule
\end{tabular}
\caption{Reduced E2 transition probabilities among the states of the second $|K|^{\pi}=2^-$ band of \ce{^{24}Mg}, corresponding to one quantum of vibrational excitation of frequency $\omega_6$. }\label{tab:EMtransitions_intraband_Band_B2u}
\end{table}

Perhaps the most timely measurements that could serve as critical tests for the $\mathcal{D}_{4h}$-symmetric G$\alpha$CM are the transition strengths for the states of the $|K|^{\pi}=2^-$ band of the present type. So far, no transitions between such states have been measured, but the G$\alpha$CM at LO indicate quite large transition E2 strengths, especially between the bandhead and the $3^-$ state at $14.019(4)~\mathrm{MeV}$ (cf. Tab.~\ref{tab:EMtransitions_intraband_Band_B2u}).

\subsection{First excited $E_g$ band}\label{sec:excited_Eg_I_band}

The $J^{\pi} = 1^+$ state at 7.7477(2) MeV has been identified as the head of the first excited $|K|^{\pi}=1^+$ band of $E_{g}$ type, corresponding to one quantum of vibrational excitation of energy $\hbar\omega_7 = 7.070(34)$ MeV, the asymmetric twisting mode ($\mathfrak{n}_7=1$).  In the large-amplitude-vibration limit, the normal mode associated with this band, $\omega_7$, enhances the ${}^{12}\mathrm{C}+{}^{12}\mathrm{C}$ cluster configuration and decay channel. 

For this vibrational mode, also the sideband with $|K|^{\pi}=3^+$ has been detected. The composition of the lowest band ($|K|^{\pi}=1^+$), agrees with Refs.~\cite{CLW93,FZB73} and partially with Ref.~\cite{GFM78} (cf. Tab.~\ref{tab:composition_excited_I_K=1+_band}). Conversely, the $K^{\pi}=3^+$ sideband starting with the $3^+$ level at 11.933(2) MeV represents an original work \cite{Ste26}. The bandhead is followed by the $4^{+}$ state at 13.910(1) MeV, the $5^{+}$ at 15.093(1) MeV, the $6^{+}$ at 16.929(3) MeV, the $(7)^{(+)}$ at 19.400(15) MeV, the $8^+$ at 22.40(2) MeV and the $9^+$ at 24.98(14) MeV. 

The interpolation of the measured energies yields a mass-specific moment of inertia about the intrinsic $x$ ($z$) axis of $126.3(39)~\mathrm{fm}^2$ ($40.4(25)~\mathrm{fm}^2$), with little modifications with respect to the ground-state band. As a consequence, the energies of the member states are described rather accurately by both the structure parameter sets in Tab.~\ref{tab:composition_excited_I_K=1+_band}. In this case, the set $(\beta_1,\beta_2) \approx (2.247,3.566)$ delivers the best results.

\begin{table}[htb!]
\centering
\begin{tabular}{ccccc}
\toprule
 \multicolumn{2}{c}{$E_{g}$ ($\omega_7$) \textsc{Band}} & Exper. [MeV] &  \multicolumn{2}{c}{G$\alpha$\textsc{cm}~LO $(\beta_1,\beta_2)$ [e $\mathrm{fm}^2$]} \\
$J^{\pi}$ & Refs. &  Ref. \cite{BaC22} &  $(2.25,3.57)$ fm & $(2.38,3.86)$ fm  \\ 
\midrule
$1^+$ & \cite{CLW93} & 7.7477(2) & 7.496 & 7.428 \\
$2^+$ & \cite{CLW93,CRJ23}   & 9.0035(2) & 8.064 & 7.937\\
$3^+$ & \cite{CLW93} & 9.45781(4) & 8.946 & 8.700 \\
$4^+$ & \cite{CLW93}  & 10.57593(8) & 10.122 & 9.717 \\
$6^+$ & \cite{CLW93} & 13.450(20) & 13.356 & 12.514 \\
$8^+$ & \cite{CLW93} & 17.190(15) & 17.766 & 16.329 \\
\bottomrule
\end{tabular}
\caption{The first $|K|^{\pi}=1^+$ band of \ce{^{24}Mg}, corresponding to one quanta of vibrational excitation of frequency $\omega_7$ ($E_{g}$ irrep). The calculated energy states refer to the parameter sets $(\beta_1,\beta_2) \approx (2.247,3.566)$ and $(2.380,3.857)~\mathrm{fm}$.  } \label{tab:composition_excited_I_K=1+_band}
\end{table}

Despite the lack of experimental measurements, a series of reduced E2 intraband transition probabilities has been calculated for the states of this band. In comparison with the electric quadrupole transition strenghts of the $A_{1g}$, $A_{2u}$, $B_{1g}$, $B_{2g}$ and $B_{2u}$ bands, the G$\alpha$CM results at LO are smaller for this band, although of the same order of magnitude (cf. Tab.~\ref{tab:EMtransitions_intraband_Band_Eg}). Theoretical counterparts from the AMD are not available yet for the states of this band. All in all, the measurement of these reduced transition probabilities are not expected to pose an experimental challenge.

\begin{table}[htb!]
\centering
\begin{tabular}{ccc}
\toprule
\multirow{2}{4.5cm}{\centering{ \textsc{Intraband} $E_{g}$ ($\omega_7$)}}  & \multicolumn{2}{c}{G$\alpha$\textsc{cm}~LO $(\beta_1,\beta_2)$ [$\mathrm{e}^2~\mathrm{fm}^{4}$]}\\
&  $(2.25,3.57)$ fm & $(2.38,3.86)$ fm  \\
\midrule
$\mathrm{B}[E2; 1^+~(7.748) \rightarrow 2^+~(9.004)]$ &166.66 & 245.71  \\
$\mathrm{B}[E2; 1^+~(7.748) \rightarrow 3^+~(9.458)]$ & 131.88 &194.86 \\
$\mathrm{B}[E2; 2^+~(9.004) \rightarrow 3^+~(9.458)]$ &  65.46 & 96.86 \\
$\mathrm{B}[E2; 2^+~(9.004) \rightarrow 4^+~(10.576)]$ & 142.33 & 210.00 \\
$\mathrm{B}[E2; 3^+~(9.458) \rightarrow 4^+~(10.576)]$ & 35.97 & 52.96 \\
$\mathrm{B}[E2; 6^+~(13.450) \rightarrow 8^+~(17.190)]$ & 137.95 & 203.53 \\
\bottomrule
\end{tabular}
\caption{Reduced E2 transition probabilities among the states of the first $|K|^{\pi}=1^+$ band of \ce{^{24}Mg}, corresponding to one quantum of vibrational excitation of frequency $\omega_7$. }\label{tab:EMtransitions_intraband_Band_Eg}
\end{table}

\subsection{First excited $E_u$ band}\label{sec:excited_Eu_I_band}

The $J^{\pi} = 1^-$ state at 8.4384(10) MeV represents the lowest level of the first excited $|K|^{\pi}=1^-$ band of $E_{u}$ type, corresponding to one quantum of vibrational excitation of energy $\hbar\omega_8 = 7.7866(27)$ MeV, the asymmetric twisting mode ($\mathfrak{n}_8=1$). The composition of this band, well documented in the literature, reflects Refs.~\cite{GFM78,CLW93} and overlaps with Ref.~\cite{KYI12,CRJ23}, except for the $4^-$ and $5^-$ states. The sideband with $|K|^{\pi}=3^-$ has been also constructed, starting from the $J^{\pi} = 3^-$ state at 12.861(3) MeV. Other member states are the $(4)^{(-)}$ state at 14.157(4) MeV, the $5^-$ at 15.214(1) MeV, the $(6)^{(-)}$ at 17.465(10) MeV, the $(7)^{(-)}$ at 19.890(15) MeV, the $(8)^{(-)}$ at 23.0(1) MeV and the $(9)^{(-)}$ at 25.4(1) MeV \cite{BaC22}.

\begin{table}[htb!]
\centering
\begin{tabular}{ccccc}
\toprule
 \multicolumn{2}{c}{$E_{u}$ ($\omega_8$) \textsc{Band}} & Exper. [MeV] &  \multicolumn{2}{c}{G$\alpha$\textsc{cm}~LO $(\beta_1,\beta_2)$ [e $\mathrm{fm}^2$]} \\
$J^{\pi}$ & Refs. &  Ref. \cite{BaC22} &  $(2.25,3.57)$ fm & $(2.38,3.86)$ fm \\ 
\midrule
$1^-$ & \cite{GFM78,CLW93,ChK20,KEO21} & 8.4384(10) & 8.192 & 8.144 \\
$2^-$ & \cite{GFM78,CLW93,KEO21} & 8.8645(2) & 8.780 & 8.653 \\
$3^-$ & \cite{CLW93,KEO21}  & 10.3336(2) & 9.662 & 9.416 \\
$5^-$ & \cite{GFM78,CLW93,KEO21} & 13.057(3) & 12.308 & 11.705 \\
\bottomrule
\end{tabular}
\caption{The first $|K|^{\pi}=1^-$ band of \ce{^{24}Mg}, corresponding to one quanta of vibrational excitation of frequency $\omega_8$ ($E_{u}$ irrep). The calculated energy states refer to the parameter sets $(\beta_1,\beta_2) \approx (2.247,3.566)$ and $(2.380,3.857)~\mathrm{fm}$.  } \label{tab:composition_excited_I_K=1-_band}
\end{table}

The interpolation of the experimental energies delivers a mass-specific moment of inertia about the intrinsic $x$ ($z$) axis of $125.3(28)~\mathrm{fm}^2$ ($42.6(21)~\mathrm{fm}^2$), overlapping the one(s) for the singly-excited $E_g$ bands within the statistical uncertainty. As previously, the $|K|^{\pi}=3^-$ sideband has been exploited for the estimation of the moment of inertia about the intrinsic $z$ axis. The G$\alpha$CM predictions at LO for the excitation energies are quite precise, with the $(\beta_1,\beta_2) \approx (2.247,3.566)$ parameters set delivering the most accurate results.

\begin{table}[htb!]
\centering
\begin{tabular}{ccc}
\toprule
\multirow{2}{4.5cm}{\centering{ \textsc{Intraband} $E_{u}$ ($\omega_8$)}}  & \multicolumn{2}{c}{G$\alpha$\textsc{cm}~LO $(\beta_1,\beta_2)$ [$\mathrm{e}^2~\mathrm{fm}^{4}$]}\\
&  $(2.25,3.57)$ fm & $(2.38,3.86)$ fm  \\
\midrule
$\mathrm{B}[E2; 1^-~(8.438) \rightarrow 2^-~(8.865)]$ &165.30 & 244.17  \\
$\mathrm{B}[E2; 2^-~(8.865) \rightarrow 3^-~(10.334)]$ &  71.38 & 104.04 \\
$\mathrm{B}[E2; 3^-~(10.334) \rightarrow 5^-~(13.057)]$ & 148.40 & 217.43 \\
\bottomrule
\end{tabular}
\caption{Reduced E2 transition probabilities among the states of the first $|K|^{\pi}=1^-$ band of \ce{^{24}Mg}, corresponding to one quantum of vibrational excitation of frequency $\omega_8$. }\label{tab:EMtransitions_intraband_Band_I_Eu}
\end{table}

Concerning the intraband EM transitions, no data from experiments is available yet. However, for completeness, certain reduced electric multipole transition probabilities have been calculated and reported in Tab.~\ref{tab:EMtransitions_intraband_Band_I_Eu}. The E2 transition strengths at LO turn out to be of the same order of magnitude of the ones of the $E_g$ band, with sizable differences in the two considered sets of structure parameters.

\subsection{Second excited $E_u$ band}\label{sec:excited_Eu_II_band}

The $J^{\pi} = 1^-$ state at 9.1462(3) MeV has been identified as the head of the second excited $|K|^{\pi}=1^-$ band of $E_{u}$ type, corresponding to one quantum of vibrational excitation of energy $\hbar\omega_9 = 8.509(14)$ MeV, the asymmetric twisting mode ($\mathfrak{n}_9=1$). This band has less evidence in the literature than the counterpart in Sec.~\ref{sec:excited_Eu_I_band} and its composition in Tab.~\ref{tab:composition_excited_II_K=1-_band} partly reflects Ref.~\cite{CLW93}.

As for the first $E_g$ mode, the structure of the $K^{\pi}=3^-$ sideband has been proposed \cite{Ste26}, although it has no roots in the literature. The latter band originates with the $J^{\pi} = 3^-$ state at 13.437(4) MeV, and is followed by the $(4)^{(-)}$ state at 14.793(1) MeV, the $5^-$ at 16.611(10) MeV, the $(6)^{(-)}$ at 18.203(10) MeV, the $(7)^{(-)}$ at 20.46(1) MeV and the $(8)^{(-)}$ at 23.19(3) MeV\cite{BaC22}. 

\begin{table}[htb!]
\centering
\begin{tabular}{ccccc}
\toprule
 \multicolumn{2}{c}{$E_{u}$ ($\omega_9$) \textsc{Band}} & Exper. [MeV] &  \multicolumn{2}{c}{G$\alpha$\textsc{cm}~LO $(\beta_1,\beta_2)$ [e $\mathrm{fm}^2$]} \\
$J^{\pi}$ & Refs. &  Ref. \cite{BaC22} &  $(2.25,3.57)$ fm & $(2.38,3.86)$ fm \\ 
\midrule
$1^-$ & \cite{CLW93} & 9.1462(3) & 8.915 & 8.867 \\
$3^-$ & \cite{CLW93} & 11.165(2) & 10.3846 & 10.138 \\
$5^-$ & \cite{CLW93}  & 13.771(3) & 13.031 & 12.427 \\
$7^-$ & New & 18.169(10) & 16.853 & 15.733 \\
\bottomrule
\end{tabular}
\caption{The second excited $|K|^{\pi}=1^-$ band of \ce{^{24}Mg}, corresponding to one quanta of vibrational excitation of frequency $\omega_9$ ($E_{u}$ irrep). The calculated energy states refer to the parameter sets $(\beta_1,\beta_2) \approx (2.247,3.566)$ and $(2.380,3.857)~\mathrm{fm}$.  } \label{tab:composition_excited_II_K=1-_band}
\end{table}

As before, the sideband has been also exploited for the estimation of the mass-specific moment of inertia about the intrinsic $x$ ($z$) axis, equal to $128.1(20)~\mathrm{fm}^2$ ($43.6(12)~\mathrm{fm}^2$), through linear regressions on the experimental excitation energies. The fitted moments of inertia are compatible with the ones of the lowest $E_g$ mode within the statistical errors, hence no substantial change in the nuclear radius and matter density distribution is expected to occur. As a consequence, the energy eigenvalues calculated at LO in the G$\alpha$CM are quite precise, as the ones in Tab.~\ref{tab:composition_excited_I_K=1-_band}. Analogously, the $(\beta_1,\beta_2) \approx (2.189,3.541)$ set of parameters provides the most accurate results (cf. Tab.~\ref{tab:composition_excited_II_K=1-_band}).

Regarding the intraband EM transitions, measurements are not available yet. Nonetheless, certain reduced electric multipole transition probabilities have been calculated and reported in Tab.~\ref{tab:EMtransitions_intraband_Band_II_Eu}. The results of the G$\alpha$CM at LO lie close to the ones of the first $E_u$ band, due to the fact that the two $|K|^{\pi}=1^-$ bands correspond to similar excitation frequencies (cf. Tab.~\ref{tab:EMtransitions_intraband_Band_I_Eu}).  

\begin{table}[htb!]
\centering
\begin{tabular}{ccc}
\toprule
\multirow{2}{4.5cm}{\centering{ \textsc{Intraband} $E_{u}$ ($\omega_9$)}}  & \multicolumn{2}{c}{G$\alpha$\textsc{cm}~LO $(\beta_1,\beta_2)$ [$\mathrm{e}^2~\mathrm{fm}^{4}$]}\\
&  $(2.25,3.57)$ fm & $(2.38,3.86)$ fm  \\
\midrule
$\mathrm{B}[E2; 1^-~(9.146) \rightarrow 3^-~(11.165)]$ &138.98 & 203.50  \\
$\mathrm{B}[E2; 3^-~(11.165) \rightarrow 5^-~(13.771)]$ & 148.90 & 218.04 \\
$\mathrm{B}[E2; 5^-~(13.771) \rightarrow 7^-~(18.169)]$ & 145.78 & 213.46 \\
\bottomrule
\end{tabular}
\caption{Reduced E2 transition probabilities among the states of the second $|K|^{\pi}=1^-$ band of \ce{^{24}Mg}, corresponding to one quantum of vibrational excitation of frequency $\omega_9$. }\label{tab:EMtransitions_intraband_Band_II_Eu}
\end{table}

\section{Conclusion}\label{sec:conclusion}

The low-lying spectrum and the electromagnetic transitions of ${}^{24}\mathrm{Mg}$ have been investigated in the framework of the geometric $\alpha$-cluster model, presented in Ref.~\cite{StS24}. In the latter, an approximation scheme for the implementation of the coupling between rotational and vibrational motion, based on the Watson Hamiltonian \cite{Wat68}, has been introduced. The tenets of the G$\alpha$CM are outlined in Ref.~\cite{Ste26}.

The square bipyramid with $\mathcal{D}_{4h}$ symmetry \cite{Bou62,HaD66} has been reproposed as an equilibrium $\alpha$-cluster configuration for ${}^{24}\mathrm{Mg}$. Due to the significant amount of spectroscopic data recorded since then, all the $9$ predicted singly-excited rotational bands have been identified \cite{Ste26}. The normal-mode vibrations of the $6\alpha$-structure permitted to bring out the connection between certain normal modes ($A_{1g}$, $A_{2u}$, $E_g$) and low-energy cluster-decay channels. Essential support to the $\mathcal{D}_{4h}$-symmetric \textit{ansatz} has been provided by the calculated M1 and E2 moments as well as the intraband and interband $\gamma$-transition strengths between the identified $\alpha$-cluster states of ${}^{24}\mathrm{Mg}$, detailed in Ref.~\cite{Ste26}. 

The perturbative application of the G$\alpha$CM at NLO \cite{Ste26} is envisaged, since the rotation-vibration coupling is responsible of the observed changes in the nuclear moments of inertia in the excited rotational bands. Additionally, the tentative inspection of doubly-excited bands, with special attention to neighbour states of decay thresholds at $13.93$ MeV (${}^{12}\mathrm{C}+{}^{12}\mathrm{C}$), 14.05 MeV ($2\alpha+{}^{16}\mathrm{O}$) and 21.21 MeV ($3\alpha+{}^{12}\mathrm{C}$) is planned.

The development of new facilities such as the \textit{variable energy gamma system} (VEGA) at ELI-NP (Măgurele, Romania), wherein photo-excitation experiments in the energy range $1$-$20$~MeV play a major role, is expected to enlarge the set of measured intraband and interband transition probabilities of ${}^{24}\mathrm{Mg}$, thus providing crucial tests for the validity of the present model.

\section*{Acknowledgements}

This proceeding is part of a greater work started under the impulse of the co-author, who carried out the first calculations in the initial phase. Indeed, K.-H. S. was deeply interested in the phenomenological aspects of clustering in nuclei and contributed to a number of experimental investigations on EM properties of light nuclei, including gyromagnetic factors \cite{SSM22}.
G.S. acknowledges funding form the \emph{Espace de Structure et de réactions Nucléaires Théorique} (ESNT) of the CEA/DSM-DAM and expresses gratitude to Serdar Elhatisari (King-Fahd University, Dhahran), Timo A. L\"ahde (FZ J\"ulich), Dean Lee (Michigan State University), Shihang Shen (Beihang University) and Vittorio Somà (CEA Paris-Saclay) for the discussions. \\

\end{document}